\documentclass[11pt,a4paper]{article}
\pdfoutput=1
\usepackage{amssymb}
\usepackage{amsmath}
\usepackage{graphicx}
\usepackage{booktabs}
\usepackage{color}
\usepackage{textcomp}
\usepackage{multirow}
\usepackage{bm}
\usepackage{caption}
\usepackage{subcaption}
\usepackage{tikz}
\usepackage{longtable}
\usepackage{colortbl}
\usepackage{jheppub}
\usepackage{dcolumn} 
\usepackage{soul}
\usepackage{mathtools}
\usepackage{float}
\usepackage{afterpage}
\usepackage[utf8]{inputenc} 

\newcommand{\cm}{\ensuremath{\mathsf{cm}}}
\newcommand{\sm}{\ensuremath{\text{-}}}

\newcommand{\pD}{\ensuremath{{D\pi}}}
\newcommand{\eD}{\ensuremath{{D\eta}}}
\newcommand{\KDs}{\ensuremath{{D_s \bar{K}}}}
\newcommand{\tx}{\checkmark}
\newcommand{\singleop}{\ensuremath{{\bar{\psi} \mathbf{\Gamma} \psi}}}
\begin{document}
\title{Coupled-Channel $D\pi$, $D\eta$ and $D_{s}\bar{K}$ Scattering from Lattice QCD}

\author[a]{Graham~Moir,} \emailAdd{graham.moir@damtp.cam.ac.uk}
\author[b]{Michael~Peardon,} \emailAdd{mjp@maths.tcd.ie}
\author[b]{Sin\'{e}ad~M.~Ryan,} \emailAdd{ryan@maths.tcd.ie}
\author[a]{Christopher~E.~Thomas,} \emailAdd{c.e.thomas@damtp.cam.ac.uk}
\author[a]{David~J.~Wilson} \emailAdd{d.j.wilson@damtp.cam.ac.uk}

\affiliation[a]{Department of Applied Mathematics and Theoretical Physics, Centre for Mathematical Sciences,
University of Cambridge, Wilberforce Road, Cambridge, CB3 0WA, UK} 

\affiliation[b]{School of Mathematics, Trinity College, Dublin 2, Ireland}

\affiliation{\vspace{0.1cm} {\rmfamily \normalsize (for the Hadron Spectrum Collaboration)}}

\abstract{We present the first lattice QCD study of coupled-channel $D\pi$, $D\eta$ and $D_{s}\bar{K}$ scattering in isospin-1/2 in three partial waves. Using distillation, we compute matrices of correlation functions with bases of operators capable of resolving both meson and meson-meson contributions to the spectrum. These correlation matrices are analysed using a variational approach to extract the finite-volume energy eigenstates. Utilising L\"uscher's method and its extensions, we constrain scattering amplitudes in $S$, $P$ and $D$-wave as a function of energy. By analytically continuing the scattering amplitudes to complex energies, we investigate the $S$-matrix singularities.  Working at $m_\pi \approx 391$ MeV, we find a pole corresponding to a $J^{P} = 0^{+}$ near-threshold bound state with a large coupling to $D\pi$. We also find a deeply bound $J^{P} = 1^{-}$ state, and evidence for a $J^{P} = 2^{+}$ narrow resonance coupled predominantly to $D\pi$. Elastic $D\pi$ scattering in the isospin-$3/2$ channel is studied and we find a weakly repulsive interaction in $S$-wave.}

\preprint{DAMTP-2016-48}

\arxivnumber{1607.07093}

\maketitle

\clearpage
\section{Introduction}\label{sec:introduction}

Over the last few years, a number of new states have been observed in both the charm-light (isospin-1/2, strangeness-0) $D$ meson and the charm-strange (isospin-0, strangeness-1) $D_s$ meson systems and experiments continue to investigate their properties and find additional states~\cite{Agashe:2014kda,Godfrey:2015dva}.  
The intermediate mass scale of the charm quark means that these systems provide a window on heavy-light dynamics away from the heavy-quark limit.
The low-lying excitations are generally in agreement with expectations from quark-potential models~\cite{Godfrey:1985} with some notable exceptions: the lightest scalar, $D_{s0}^{\star}(2317)$, and axial vector, $D_{s1}(2460)$, charm-strange mesons were expected to be broad and above the relevant strong-decay threshold ($D K$ and $D^{\star} K$ respectively), but they were both observed to be narrow and below threshold. A number of possible explanations have been put forward~\cite{Godfrey:2015dva,Prencipe:2015kva}. On the other hand, the corresponding charm-light mesons, $D_{0}^{\star}(2400)$ and $D_{1}(2430)$, were both observed, as expected, to be broad resonances. 
The dynamics in the charm-light and charm-strange sectors are expected to be similar but the different masses that enter and the relative position of thresholds appear to be playing an important role. In this study, we investigate charm-light states as a step towards understanding these differences.

Lattice QCD provides a method for performing first-principles non-perturbative computations of the masses and other properties of hadrons within QCD.  Correlation functions are computed numerically by Monte-Carlo sampling gauge configurations on a discretised Euclidean spacetime of finite volume, yielding a discrete spectrum of energy eigenstates.  One virtue of lattice QCD is that it is systematically-improvable, permitting increasingly precise spectra to be obtained through efficient use of computational resources.  

In recent years there has been significant progress in computing spectra of excited hadrons using lattice QCD.  In particular, the L\"{u}scher method and its extensions for relating finite-volume spectra to scattering amplitudes are now well established for elastic~\cite{Luscher:1990ux,Luscher:1991cf, Rummukainen:1995vs, Feng:2004ua, Christ:2005gi, Kim:2005gf, Bernard:2008ax, Bernard:2010fp, Leskovec:2012gb, Gockeler:2012yj} and coupled-channel \cite{He:2005ey, Hansen:2012tf, Briceno:2012yi, Guo:2012hv, Briceno:2014oea} hadron-hadron scattering.  These methods have been demonstrated in a number of applications, notably for the $\rho$-resonance seen in $P$-wave $\pi\pi$ scattering~\cite{Aoki:2007rd, Feng:2010es, Lang:2011mn, Aoki:2011yj, Pelissier:2012pi, Dudek:2012xn, Wilson:2015dqa, Bali:2015gji, Bulava:2016mks, Guo:2016zos}, and for the $\sigma$ resonance seen in $S$-wave $\pi\pi$ scattering~\cite{Briceno:2016mjc}. It has also recently been shown that, with sufficiently extensive and precise spectra, information on coupled-channel hadron-hadron scattering amplitudes can be obtained~\cite{Dudek:2014qha,Wilson:2014cna,Wilson:2015dqa,Dudek:2016cru} -- this is crucial for understanding highly excited states that are typically kinematically permitted to decay into several channels.

Recent lattice QCD investigations of charm-light mesons beyond the lightest pseudoscalar and vector include Refs.~\cite{Moir:2013ub, Bali:2015lka, Kalinowski:2015bwa, Cichy:2016bci}, but these calculations were not sensitive to meson-meson energy levels and so could not robustly determine states close to threshold or properly take into account the resonant nature of states above threshold.
Elastic $D \pi$ scattering was investigated to a limited extent in Ref.~\cite{Mohler:2012na}, and in Ref.\ \cite{Liu:2012zya} the isospin-3/2 $D \pi$ scattering length was computed and used as an input to a chiral unitary approach to indirectly calculate the isospin-1/2 $D \pi$ scattering length.

Here we present a lattice QCD investigation of isospin-1/2 coupled-channel $D \pi$, $D \eta$, $D_s \bar{K}$ scattering relevant for charm-light mesons, the first coupled-channel calculation using \emph{ab-initio} methods in the charm sector: the $D\pi$ channel opens first with $D\eta$ and $D_{s}\bar{K}$ found close together a little higher in energy. We compute precise finite-volume spectra in many different symmetry channels for various momenta on multiple lattice volumes.  From these spectra, we use extensions of the L\"{u}scher method to determine infinite-volume scattering amplitudes.  Considering coupled channels enables us to constrain the amplitudes over a larger range in energy than would be possible with elastic scattering; the extensive spectra allow us to determine these amplitudes robustly and to constrain the effect of higher partial waves.
We also study elastic $D \pi$ scattering in the exotic-flavour isospin-$3/2$ channel for which preliminary results have already appeared \cite{Moir:2013yfa}.  

The remainder of this paper is laid out as follows: in Section~\ref{sec:calc_details} we give a brief description
of the lattice ensembles used in this work, along with an overview of our methodology for extracting finite-volume
spectra from two-point correlation functions. We then discuss how we obtain scattering amplitudes from finite-volume spectra. In Section~\ref{sec:iso_onehalf} results for isospin-$1/2$ coupled-channel $D\pi$, $D\eta$, $D_{s}\bar{K}$ scattering are presented and in Section~\ref{sec:iso_three_half} we show our isospin-$3/2$ $D\pi$ results.  We summarise in Section~\ref{sec:conclusions}.

%%%%%%%%%%%%%%%%%%%%%%%%%%%%%%%%%%%%%%%%%%%%%%%%%%%%%%%%%%%%%%%%%%%%%%%%%%%%%
\section{Calculation Details}\label{sec:calc_details}

\begin{table}[t]
\begin{center}
\begin{tabular}{c|c|c|c}
$(L/a_s)^3\times T/a_t$ & $N_{\rm cfgs}$ & $N_{\rm tsrcs}$ & $N_{\rm vecs}$ \\
\hline
$16^{3}\times 128$ & 479 & 4   & 64  \\
$20^{3}\times 128$ & 603 & 3   & 128 \\
$24^{3}\times 128$ & 553 & 2-3 & 162 \\
\end{tabular}
\caption{The gauge field ensembles used in this study. The volume is given by,
$(L/a_s)^3 \times (T/a_t)$, where $L$ and $T$ are respectively the spatial and temporal extents of the
lattice. The number of gauge field configurations used, $N_{\rm cfgs}$, and the number
of time-sources used per configuration, $N_{\rm tsrcs}$, are shown. $N_{\rm vecs}$ refers to
the number of eigenvectors used in the distillation framework.}
\label{tab:ensembles}
\end{center}
\end{table}

We use an anisotropic lattice formulation where the temporal lattice spacing, $a_t$, is finer than the spatial lattice spacing, $a_s \approx 0.12$ fm, with $\xi \equiv a_s/a_t \approx 3.5$. The finer temporal resolution is crucial in allowing us to accurately resolve the time dependence of two-point correlation functions enabling a precise determination of finite-volume energies.
The gauge sector is described by a tree-level Symanzik-improved anisotropic action while in the fermionic sector a tadpole-improved anisotropic Sheikholeslami-Wohlert (clover) action, with stout-smeared gauge fields and $N_{f}=2+1$ flavours of dynamical quarks, is used.  For these ensembles, $m_{\pi} \approx 391$ MeV, while the heavier dynamical quark is tuned to approximate the physical strange quark. The three different spatial volumes used are summarised in Table~\ref{tab:ensembles}.  Full details of the formulation are given in Refs.~\cite{Edwards:2008ja, Lin:2008pr}.

The same action is used for valence charm quarks as for the light and strange quarks (with tadpole-improved tree-level clover coefficients), where the charm quark mass parameter has previously been tuned using the physical $\eta_{c}$ mass \cite{Liu:2012ze}. By fitting to a relativistic dispersion relation, the anisotropy measured from the $D$-meson has been determined to be $\xi_{D} = 3.454(6)$ \cite{Moir:2013ub}, which agrees with the value measured from the pion $\xi_{\pi} = 3.444(6)$ \cite{Dudek:2012gj}. In this work, we use $\xi_{\pi}$ as the anisotropy and present our determination of scattering amplitudes incorporating its statistical uncertainty.

When we quote values in physical units, we set the scale by comparing the $\Omega$-baryon mass determined on these ensembles, $a_t m_\Omega = 0.2951$~\cite{Edwards:2011}, to the physical mass, $m_\Omega^{\textrm{phys}}$, via $a^{-1}_t = \frac{m_\Omega^{\textrm{phys}}}{a_t m_\Omega}$, leading to $a^{-1}_t = 5.667$ GeV.

%%%%%%%%%%%%%%%%%%%%%%%%%%%%%%%%%%%%%%%%%%%%%%%%%%%%%%%%%%%%%%%%%%%%%%%%%%%%%
\subsection{Finite-Volume Spectra}
\label{subsec:spectral_extraction}

To determine the discrete spectrum of finite-volume energies we compute Euclidean two-point correlation functions,
\begin{equation}
C(t) = \langle 0|{\cal O}(t){\cal O}^{\dagger}(0)|0\rangle~,
\end{equation}
where the interpolating operators, ${\cal O}^{\dagger}$ and ${\cal O}$, are chosen to have the quantum numbers of the states of interest.  
In order to robustly extract many energy levels in each channel we follow our well established procedure~\cite{Dudek:2010}: a matrix of two-point correlation functions, $C_{ij}(t)$, is computed using a basis of operators with the relevant quantum numbers.  A variational procedure \cite{Michael:1985} is employed, which amounts to solving a generalised eigenvalue problem,
\begin{equation}
C_{ij}(t) v^{\mathfrak{n}}_j = \lambda_{\mathfrak{n}}(t,t_0) C_{ij}(t_0) v^{\mathfrak{n}}_j~.
\end{equation}
\sloppy The energies then follow from analysing the time dependence of the eigenvalues (known as principal correlators), $\lambda_{\mathfrak{n}}(t,t_0)$.  We fit each principal correlator to the form ${(1 - A_{\mathfrak{n}}) e^{-E_{\mathfrak{n}} (t-t_{0})} + A_{\mathfrak{n}} e^{-E'_{\mathfrak{n}} (t-t_{0})}}$, where the fit parameters are $E_{\mathfrak{n}}$, $A_{\mathfrak{n}}$ and $E'_{\mathfrak{n}}$; the second exponential proves useful in stabilising the fit by ``mopping up'' excited state contamination. The eigenvectors, $v^{\mathfrak{n}}_{j}$, are related to the operator-state overlaps, $Z_i^{\mathfrak{n}} \equiv \langle \mathfrak{n} | \mathcal{O}_i^\dagger | 0 \rangle$, and also give weights for constructing a variationally-optimal operator to interpolate state $\mathfrak{n}$, $\Omega_\mathfrak{n}^\dagger \sim \sum_i v_i^\mathfrak{n} \mathcal{O}_i^\dagger$.

Working in a finite cubic volume with periodic boundary conditions quantises the allowed momenta, $\vec{P} = \frac{2 \pi}{L}(n_x, n_y, n_z)$, where $(n_x, n_y, n_z)$ is a triplet of integers.  We will use a shorthand notation when labelling momenta in which we omit the $\frac{2 \pi}{L}$ factor, e.g.\ $\vec{P} = [n_x, n_y, n_z]$ or $[n_x n_y n_z]$.  The finite lattice volume also breaks the rotational symmetry of an infinite-volume continuum: for mesons at rest the relevant symmetry is that of a cube, the octahedral group with parity $\mathrm{O}_{h}$, whereas for mesons at non-zero momentum, $\vec{P}$, the symmetry is reduced further to that of the little group, $\mathrm{LG}(\vec{P})$~\cite{Moore:2005dw}.  As a result, the continuum spin, $J$, is no longer a good quantum number and instead states must be labelled by the irreducible representations (\emph{irreps}) of $\mathrm{O}_{h}$ or $\mathrm{LG}(\vec{P})$. The consequences of this for scattering will be discussed below in Section~\ref{subsec:scattering_analysis}.

To reliably extract the many energy levels required to map out scattering amplitudes, we compute $C_{ij}(t)$ for large bases of interpolating operators with various structures. These include fermion-bilinear $\bar{q}{q}$ operators, $\bar{\psi}\Gamma \overleftrightarrow{D} \dots \psi$~\cite{Dudek:2010,Thomas:2011rh}, as well as those resembling the combination of two-mesons, $\sum_{\vec{p_1}, \vec{p_{2}}} \mathcal{C}(\vec{p_{1}}, \vec{p_{2}}) \Omega_{M_1}^{\dagger}(\vec{p_{1}}) \Omega_{M_2}^{\dagger}(\vec{p_{2}})$~\cite{Dudek:2012gj,Dudek:2012xn}, where $\Omega_{M_i}(\vec{p_i})$ is a variationally-optimal linear combination of fermion-bilinear operators to interpolate meson $M_i$ with momentum $\vec{p_i}$, and $\mathcal{C}$ is a generalised Clebsch Gordan coefficient.  For isospin-$1/2$ scattering we use $\bar{q}{q}$ operators along with $D \pi$, $D \eta$ and $D_s \bar{K}$ ``two-meson'' operators.  For isospin-$3/2$ we only use $D \pi$ operators -- there are no $\bar{q}{q}$ operators with this isospin. The operator bases we use are listed in Tables~\ref{tab_ops_A1}, \ref{tab_ops_P}, \ref{tab_ops_D}, \ref{tab_I32_ops_A1} and \ref{tab_I32_ops_P} in Appendix \ref{app:op_tables}. 

We use the distillation framework~\cite{Peardon:2009} which enables us to efficiently compute correlation functions involving operators with various structures where each operator is projected onto a definite momentum.  In Table~\ref{tab:ensembles} we give the number of distillation vectors, $N_{\rm vecs}$, used on each lattice volume, along with the number of time-sources used per configuration, $N_{\rm tsrcs}$.

\begin{figure}[t]
\begin{center}
\includegraphics[width=0.99\textwidth]{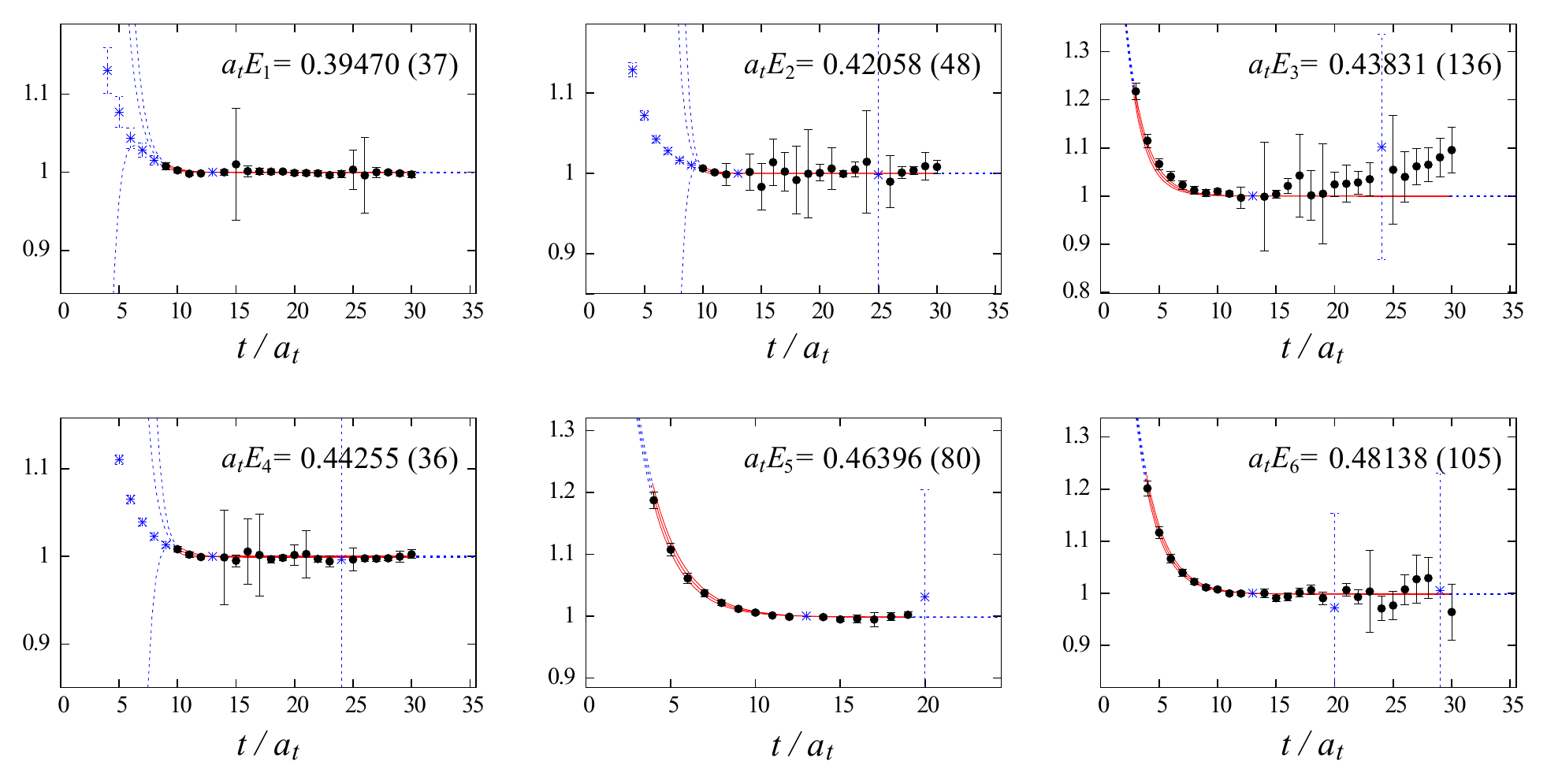}
\caption{Principal correlators, labelled by the extracted energy, determined on the $20^3$ volume in the $[000]A_1^+$ irrep with isospin-$1/2$. Points show $\lambda_{\mathfrak{n}}(t,t_{0} = 13)$ and error bars correspond to the one-sigma statistical uncertainty. In each plot the dominant time-dependence, $e^{-E_{\mathfrak{n}}(t-t_0)}$, has been divided out so that a horizontal line is observed when a single exponential dominates. Curves show fits to the form described in the text; the red curves show the fit range and blue points are not included in the fit.}
\label{fig:prin_corrs}
\end{center}
\end{figure}

As an example of the quality of the signals extracted, in Fig.~\ref{fig:prin_corrs} we show the principal correlators from the $[\vec{P}] \Lambda^P = [000]A_1^+$ irrep with isospin-$1/2$ on the $20^3$ volume; the resulting spectrum is shown in the leftmost panel of Fig.~\ref{spec_A1}.  In each plot, the dominant time-dependence, $e^{-E_{\mathfrak{n}}(t-t_0)}$, has been divided out and we observe a horizontal line when a single exponential dominates.  The principal correlators shown here are representative of all those determined within our calculation.

%%%%%%%%%%%%%%%%%%%%%%%%%%%%%%%%%%%%%%%%%%%%%%%%%%%%%%%%%%%%%%%%%%%%%%%%%%%%%
\subsection{Scattering Amplitudes from Finite-Volume Spectra}
\label{subsec:scattering_analysis}

Having determined the finite-volume spectra, we relate these to infinite-volume scattering amplitudes using the L\"{u}scher method~\cite{Luscher:1990ux,Luscher:1991cf} and its extensions to moving frames~\cite{Rummukainen:1995vs,Kim:2005gf,Fu:2011xz,Leskovec:2012gb} and coupled-channels~\cite{Liu:2005kr,Hansen:2012tf,Briceno:2012yi,Guo:2012hv}.  In this approach, the dependence of the spectra on finite volume is used as a tool but exponentially-suppressed corrections in the volume are neglected -- typically the leading such corrections fall off as $e^{-m_\pi L}$ and, since our volumes have $m_\pi L \sim 4$ to 6, we can safely neglect these.  

For lattice irrep $\Lambda$ and overall momentum $\vec{P} = \frac{2 \pi}{L} \vec{d}$, the relation between the finite-volume spectra and the infinite-volume scattering $t$-matrix is given by
\begin{equation}
\det\Big[ \delta_{ij}\delta_{\ell\ell^\prime}\delta_{nn^\prime} + i\rho^{}_i\, t^{(\ell)}_{i j}
\left(\delta_{\ell\ell^\prime} \delta_{nn^\prime} + i \mathcal{M}^{\vec{d},\Lambda}_{\ell n;\ell^\prime n^\prime}(q^2_i) \right)
\Big] = 0 \, ,
\label{eqn:luscher}
\end{equation}
where $i$ and $j$ label the scattering channels, $\rho_{i} = 2k_{i}/E_{\cm}$ is the phase-space factor for channel $i$, $k_{i}$ denotes the the momentum in the centre of momentum frame, and $t^{(\ell)}_{i j}$ is the infinite-volume $t$-matrix for partial wave $\ell$.  $\mathcal{M}(q^2_i)$ is a matrix of known functions of
$q_i^{2} = \left(k_{i}L/{2\pi}\right)^{2}$ where $\ell$ and $\ell'$ are partial waves that can subduce into irrep $\Lambda$ and the index $n$ indicates the $n$'th subduction of partial wave $\ell$ (similarly for $n'$ and $\ell'$) -- we show the pattern of subductions in Table~\ref{tab:subductions} in Appendix~\ref{app:tables}.  The mixing between different partial waves, encoded in $\mathcal{M}$, is a consequence of the reduced symmetry of the finite cubic volume.  From the table it is clear that even and odd partial waves mix when the overall momentum is non-zero (this is a consequence of the unequal mass of the scattering mesons).

Given an infinite-volume $t$-matrix, Eq.~\ref{eqn:luscher} can be solved to find the finite-volume spectrum, $\{E_{\cm}\}$.  However, the reverse is in general an under-constrained problem: for $N$ coupled channels there is one equation for each energy level but $(N^{2}+N)/2$ energy-dependent parameters in the $t$-matrix.  Hence, even neglecting the mixing between different partial waves, there is insufficient information to determine the $t$-matrix directly.
In order to circumvent this difficulty, we follow Ref.~\cite{Wilson:2014cna} and parametrise the energy-dependence of the $t$-matrix with a relatively small number of parameters.  Using Eq.~\ref{eqn:luscher}, this parametrised $t$-matrix gives a spectrum $\{E_\cm^{\textrm{par}}\}$ and we vary the parameters to fit $\{E_\cm^{\textrm{par}}\}$ to our computed spectrum $\{E_\cm\}$, minimising the $\chi^2$ function defined in Eq.~8 in \cite{Wilson:2014cna}. By analytically continuing the resulting $t$-matrix into the complex $s = E^{2}_{\cm}$ plane, we can determine the pole and residue content of the scattering amplitude, which are arguably the least method-dependent quantities that can be compared between analyses. We consider a wide range of different parametrisations to ensure the final answer does not depend on a particular form used.

When considering elastic scattering, the $t$-matrix can be described by a single energy-dependent parameter, the scattering phase shift, $\delta_\ell(E_\cm)$, where $t^{(\ell)} = \tfrac{1}{\rho} e^{i \delta_\ell} \sin \delta_\ell$.  
Two commonly used parametrisations in this case are the effective range expansion and the relativistic Breit-Wigner. The first is given by
\begin{equation}
k_i^{2\ell+1}\cot\delta_\ell = \frac{1}{a_\ell} + \frac{1}{2}r_\ell k^2_i + P_2 k^4_i
+\mathcal{O}\!\left(k_i^6\right),
\label{eqn:effective_range}
\end{equation}
where the constants $a_{\ell}$ and $r_{\ell}$ are known as the scattering length and the effective range
respectively. The second, which is commonly used to describe a resonance, is given by
\begin{equation}
t^{(\ell)}(s) =\frac{1}{\rho(s)} \frac{ \sqrt{s}\, \Gamma_\ell(s)}{m_R^2-s-i \sqrt{s}\,  \Gamma_\ell(s)},
\label{eq_bw}
\end{equation} 
where $m_{R}$ is known as the Breit-Wigner mass. $\Gamma_\ell(s)$ is the energy-dependent width which can
be parametrised in terms of the coupling $g_{R}$,
$\Gamma_\ell(s)=\frac{g_{R}^2}{6\pi}\frac{k^{2\ell+1}}{s\, m_R^{2\left(\ell-1\right)}}$, ensuring the
correct near-threshold behaviour.

For coupled-channel scattering, the relationship between the $t$-matrix, phases and inelasticities becomes more complicated. For each channel $i$, the phases, $\delta_{i}$, can be extracted from the diagonal elements of the $t$-matrix for each partial wave $\ell$,
\begin{equation}
t_{ii}= \frac{\eta_{i} e^{2i\delta_{i}} - 1 }{2i\rho_{i}} \, ,
\label{eq_phases_3x3}
\end{equation}
which also provides a convention for determining the inelasticities $\eta_{i}$. When parametrising the $t$-matrix in the coupled-channel case, we make use of the $K$-matrix formalism, where the inverse of the $t$-matrix for a partial wave $\ell$ is given by
\begin{equation}
t^{-1}_{ij}(s) = \frac{1}{(2k_i)^\ell} K^{-1}_{ij}(s) \frac{1}{(2k_j)^\ell} + I_{ij}(s)~.
\label{eqn:t_matrix_param}
\end{equation}
The factors $(2k_{i})^{-\ell}$ ensure correct behaviour in the proximity of kinematic thresholds \cite{Guo:2010gx}, while we parametrise the symmetric matrix $K^{-1}_{ij}(s)$. There is of course some freedom in this parametrisation and in this work we will make use a variety of forms which can be written generally as 
\begin{equation}
K_{ij}=\left(g_i^{(0)} + g_i^{(1)}s\right) \left(g_j^{(0)} + g_j^{(1)}s\right) \frac{1}{m^2 - s } + \gamma_{ij}^{(0)} + \gamma_{ij}^{(1)}s~,
\label{eq_K_general}
\end{equation}
where $g_i^{(n)}$ and $\gamma_{ij}^{(n)}$ are real free parameters. Unitarity of the $t$-matrix is ensured when $K^{-1}_{ij}(s)$ is real for real values of $s$, Im[$I_{ij} (s)$] $= -\delta_{ij} \rho_{i}(s)$ for energies above the kinematic threshold of channel $i$ and Im[$I_{ij} (s)$] $= 0$ for energies below that same threshold. Since unitarity does not directly constrain Re[$I_{ij}(s)$], there is some
freedom in its choice, with the simplest option being being $\mathrm{Re}[I_{ij}(s)]=0$, i.e.\, $I_{ij}(s)=-i\rho_{ij}(s)$. A different choice is the Chew-Mandelstam prescription~\cite{Chew:1960iv}, which uses the known imaginary part of $I_{ij}(s)$ to determine the real contribution through a dispersion relation. In this scheme, which captures many of the correct analytic properties of scattering amplitudes, the dispersion integral is made finite by subtraction at an arbitrary point. For elastic $S$-wave $D\pi$ scattering we subtract at threshold and in all other cases we subtract at the $K$-matrix pole parameter ($s = m^2$).\footnote{In some cases one subtraction point leads to significantly smaller correlation between the parameters.}
Details of our implementation are given in Ref.~\cite{Wilson:2014cna}. In this work, we will only consider energies far from the left-hand cut (which arises from $t$-channel exchanges). As a consequence, we do not consider the effects of this cut.

\section{Results: Isospin-$1/2$}\label{sec:iso_onehalf}

We now present the results of our calculations in the isospin-$1/2$ channel. First we discuss the finite-volume spectra obtained from the variational procedure, before moving on to discuss both elastic $D\pi$ and coupled-channel $D\pi$, $D\eta$ and $D_{s}\bar{K}$ scattering. We end the section with an interpretation of our results in terms of poles of the infinite-volume scattering matrix.

\subsection{Finite-Volume Spectra}

\begin{table}[t!]
\begin{center}
\begin{tabular}{c|c}
meson & $a_t m$ \\
\hline
$\pi$         & 0.06906(13) \\
$K$           & 0.09698(9)\\
$\eta$        & 0.10364(19)\\ 
$\eta^\prime$ & 0.1641(10)\\
$\omega$      & 0.15678(41)\\
$D$           & 0.33265(7)\\
$D_s$         & 0.34426(6)\\
$D^\star$     & 0.35415(17)\\
$D_s^\star$   & 0.36508(88)
\end{tabular}
\hspace{1cm}
\begin{tabular}{c|c}
channel & $a_t E_\mathrm{thr}$ \\
\hline
$D\pi$            & 0.40171(15)\\
$D^{\star}\pi$      & {\it 0.42321(21)}\\
$D\eta$           & 0.43629(20)\\
$D_{s}\bar{K}$       & 0.44124(11)\\
$D^{\star} \eta$& {\it 0.45779(21)}\\
$D^{\star}_{s}\bar{K}$& {\it 0.46206(88)}\\
$D\pi\pi$         & {\it 0.47077(27)}\\
$D\omega$         & {\it 0.48943(42)}\\
$D^{\star}\pi\pi$   & 0.49227(31) 
\end{tabular}
\caption{Relevant stable meson masses and kinematic thresholds on our ensembles \cite{Dudek:2012gj,Liu:2012ze,Moir:2013ub,Dudek:2016cru}. Those shown in italics do not contribute to pseudoscalar-pseudoscalar scattering in $S$-wave.}
\label{tab_massthresholds}
\end{center}
\end{table}

\begin{figure}[t!]
\begin{center}
\includegraphics[width=0.99\textwidth]{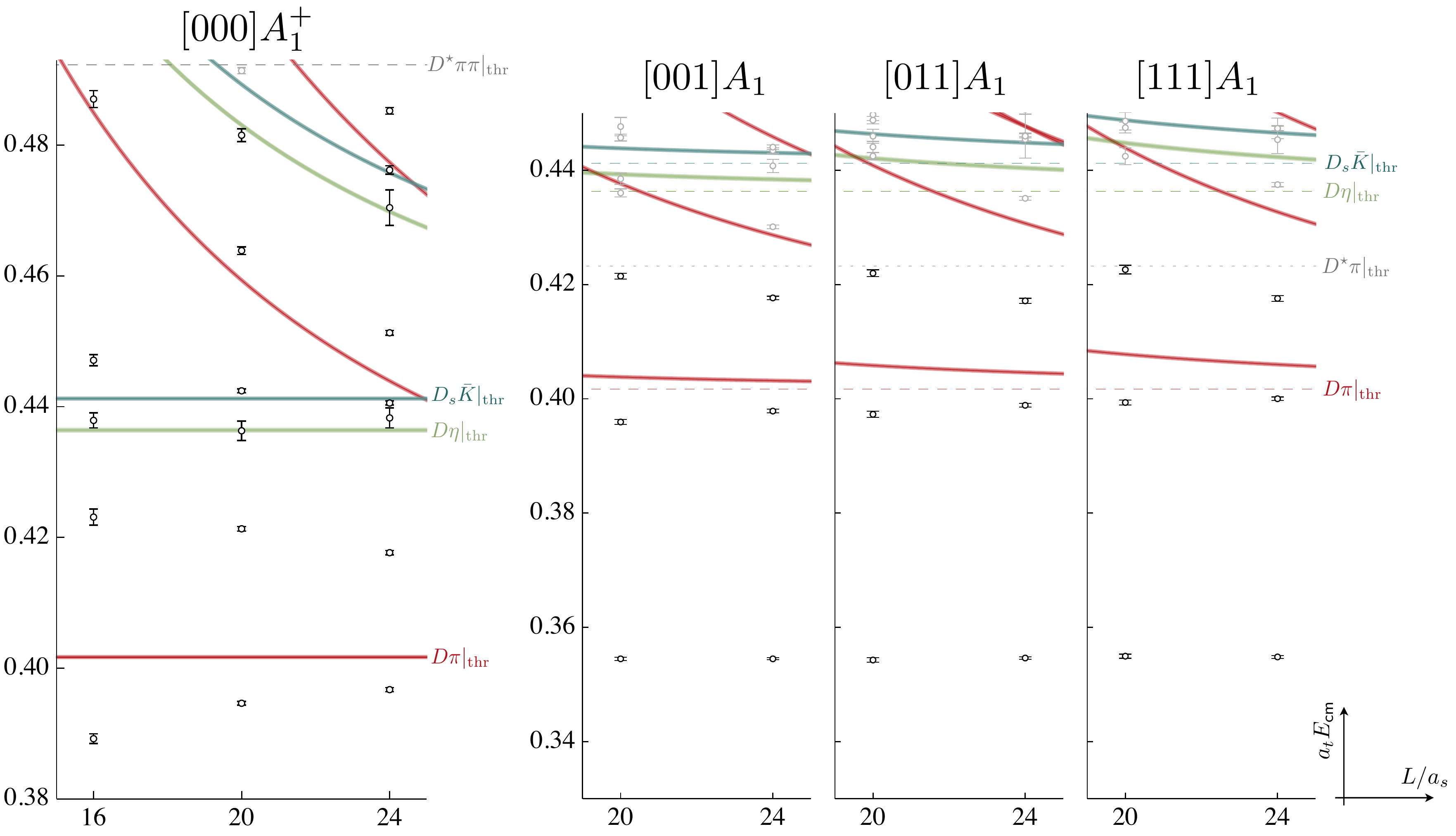}
\caption{Finite-volume spectra with isospin-$1/2$ obtained in the $[000]A_{1}^{+}$, $[001]A_1$, $[011]A_1$ and $[111]A_1$ irreps. Points are the energy levels determined using the interpolating operators listed in Table~\ref{tab_ops_A1} (with error bars showing the statistical uncertainty); those coloured black are used in our subsequent scattering analysis whereas those coloured grey are not.
Solid curves represent non-interacting energies while dashed lines correspond to channel thresholds. The colour coding is as follows: red corresponds to $D\pi$, green to $D\eta$ and blue to $D_{s}\bar{K}$. The grey dashed and dotted lines show the opening of channels for which we have not included operators in our variational procedure.}
\label{spec_A1}
\end{center}
\end{figure}

\begin{figure}[t!]
\begin{center}
\includegraphics[width=0.99\textwidth]{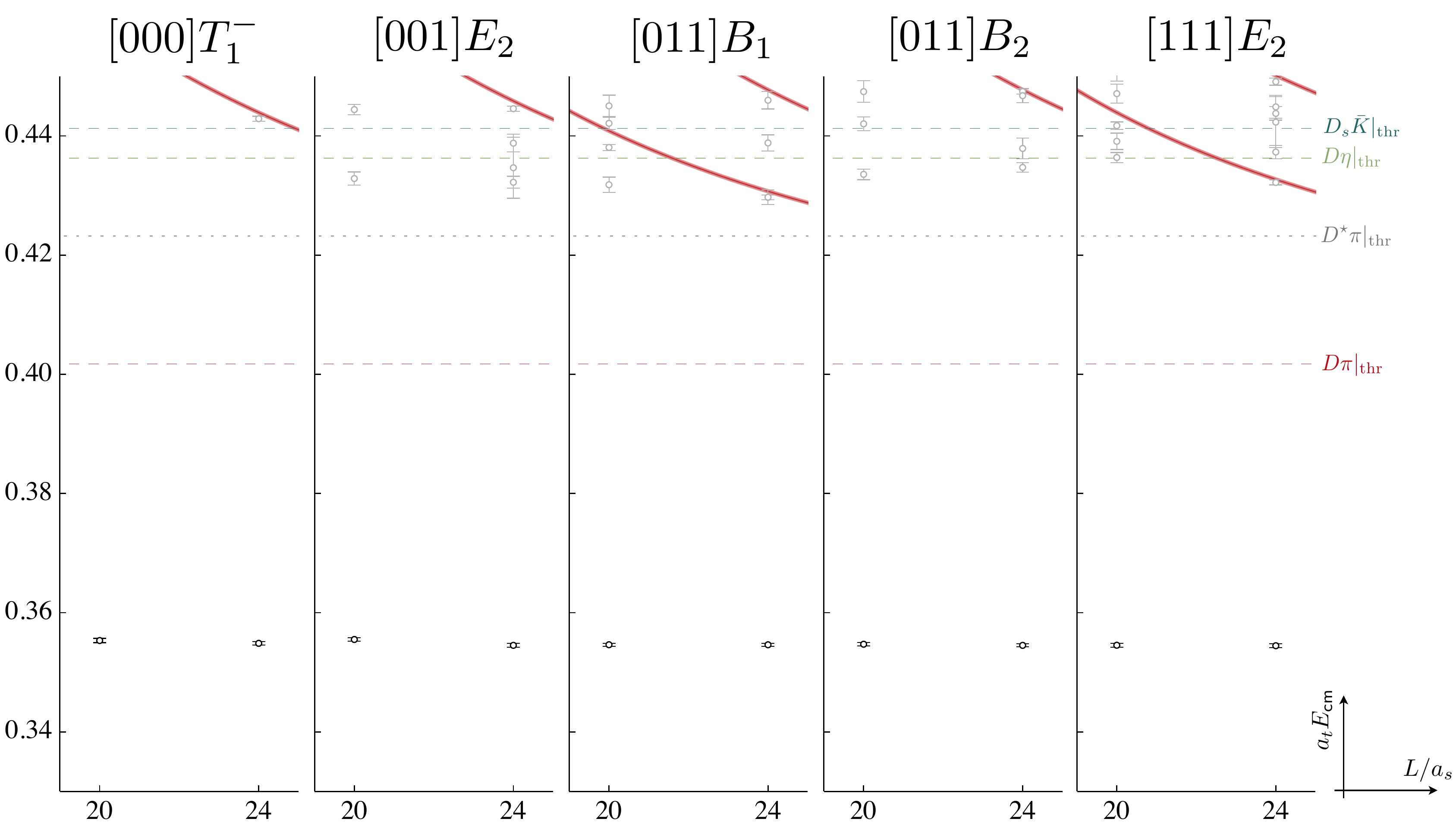}
\caption{As Fig.~\ref{spec_A1} but for the $[000]T_{1}^{-}$, $[001]E_{2}$, $[011]B_{1}$, $[011]B_{2}$ and
$[111]E_{2}$ irreps. The interpolating operators used are listed in Table \ref{tab_ops_P}.}
\label{spec_P}
\end{center}
\end{figure}

\begin{figure}[t!]
\begin{center}
\includegraphics[width=0.80\textwidth]{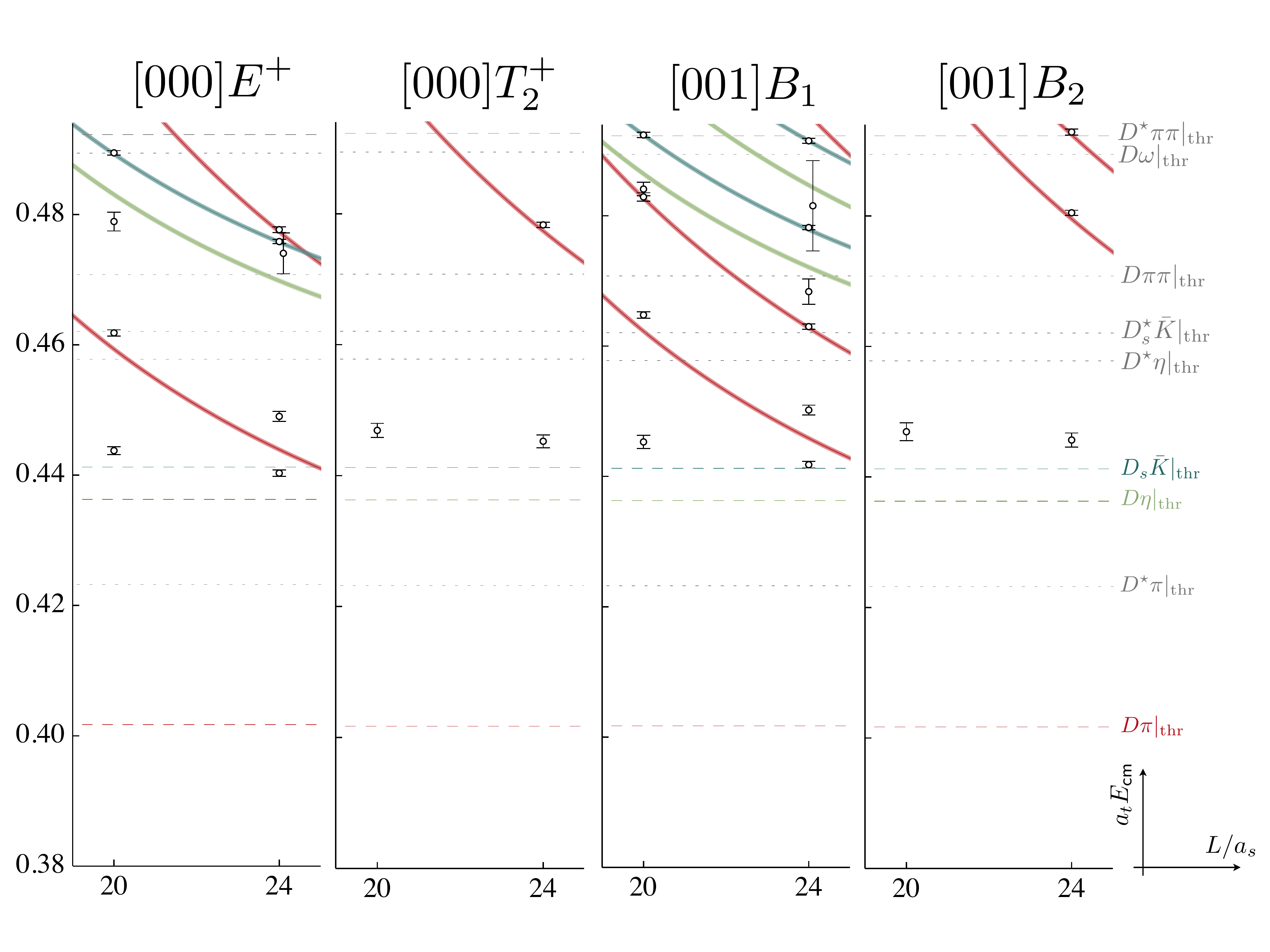}
\caption{As Fig.~\ref{spec_A1} but for the $[000]E^{+}$, $[000]T_{2}^{+}$, $[001]B_{1}$ and $[001]B_{2}$ irreps.  The operators used are listed in Table \ref{tab_ops_D}.}
\label{spec_D}
\end{center}
\end{figure}

Following the procedure described in Section \ref{subsec:spectral_extraction}, the large bases of interpolating operators listed in Tables~\ref{tab_ops_A1}, \ref{tab_ops_P} and \ref{tab_ops_D} in Appendix \ref{app:op_tables} are used to determine finite-volume spectra in a number of lattice irreps, $[\vec{P}]\Lambda^{(P)}$, where parity $P$ is only a good quantum number when the system has zero overall momentum. These are shown in Figs.~\ref{spec_A1},~\ref{spec_P} and \ref{spec_D}, where the black and grey points correspond to the extracted finite-volume energy levels; only black points are used in the subsequent scattering analyses. The red, green and blue curves (dashing) are the non-interacting $D\pi$, $D\eta$ and $D_{s}\bar{K}$ energies (thresholds) respectively. The grey dotted lines show the opening of channels for which we have not included interpolating operators in our variational procedure. Relevant meson masses and multi-meson thresholds are given in Table~\ref{tab_massthresholds}.

The leftmost panel of Fig.~\ref{spec_A1} shows the spectrum obtained in the $[000]A_1^{+}$ irrep, whose lowest partial wave contribution comes from $\ell = 0$. The lowest energy level, which has a large overlap with our $D_{000}\pi_{000}$ operator (defined in Table \ref{tab_ops_A1}), shows a volume dependent shift away from the $D\pi$ threshold. Furthermore, there appears to be an ``extra" energy level compared to the number of non-interacting multi-meson levels expected in the energy region below the $D\eta$ threshold. These features may point to a non-trivial meson-meson interaction in $S$-wave within the energy region around $D\pi$ threshold. In Fig.~\ref{spec_P}, we show spectra extracted in irreps whose lowest partial wave contribution comes from $\ell = 1$. In this case we observe a level far below the $D\pi$ threshold at $a_{t}E_{\cm} \approx 0.35$, suggesting that a vector bound state is present. In Fig.~\ref{spec_A1}, we also show the $[\vec{P} \neq 0]A_{1}$ irreps where both $S$ and $P$-waves can contribute. Here we find further evidence for a non-trivial $S$-wave interaction near the $D\pi$ threshold and the deeply bound vector state; we observe volume dependent shifts near the $D\pi$ threshold along with the appearance of an ``extra" energy level, while also finding an energy level far below the $D\pi$ threshold at $a_{t}E_{\cm} \approx 0.35$.

The spectra shown in Fig.~\ref{spec_D} have $\ell = 2$ as their lowest contributing partial wave. Within the energy range $0.44 \lesssim a_{t}E_{cm} \lesssim 0.46$, we observe significant shifts of the energy levels away from non-interacting energies along with the presence of an ``extra" energy level, indicative of non-trivial interactions in $D$-wave.

\subsection{Elastic $D\pi$ Scattering}\label{subsec:elastic_Dpi}

To begin our scattering analysis we consider the region where only the $D\pi$ channel is kinematically open, that is below the $D\eta$ threshold when the system has zero overall momentum and below the $D^{\star}\pi$ threshold when the system has overall non-zero momentum. In the near-threshold region, higher partial waves are suppressed in proportion to $k^{2\ell+1}$ in the absence of resonances. This is important because partial waves can mix; for $D\pi$ scattering in a finite volume, the allowed partial waves are $\ell^{P}=0^{+},1^{-},2^{+},...$, where parity, $P$, is only a good quantum number when the system has overall zero momentum.

\subsubsection{$S$-wave}

By following the procedure described in Section \ref{subsec:scattering_analysis}, we determine the $S$-wave scattering amplitude using only the $[000]A_{1}^{+}$ irrep, neglecting $\ell \geq 4$ contributions. Taking the lowest
two levels in each volume and parametrising the $t$-matrix using a $K$-matrix containing a pole term and a constant, the $\chi^{2}$ function is minimised, obtaining\\
\begin{center}
\begin{tabular}{rll}
$m \;\; =$                         & $(0.396 \pm 0.003 \pm 0.002) \cdot a_t^{-1}$   & 
\multirow{3}{*}{ $\begin{bmatrix} 1 \quad & -0.72 &  -0.39  \\ 
                                     &  1 \quad   &   0.91  \\
                                     &            & 1 \quad  \end{bmatrix}$ } \\
$g \;\; =$                 & $(0.65 \pm 0.09 \pm 0.07) \cdot a_t^{-1}$   & \\
$\gamma \;\; =$   & $\,\,\,\;15 \pm 4 \pm 2 $   & \\[1.3ex]
&\multicolumn{2}{l}{ $\chi^2/ N_\mathrm{dof} = \frac{3.43}{6-3} = 1.14 $\,.}
\end{tabular}
\end{center}
\vspace{-1cm}
\begin{equation} \label{rest_fit_A1}\end{equation}\\
For each parameter, our convention is that the first uncertainty reflects the $\chi^2$ minimisation and the second uncertainty is obtained by varying the scattered meson masses and the anisotropy within their statistical uncertainties. The matrix shows the correlation between each parameter. Although this form does not demand the existence of a nearby pole in the $t$-matrix, it permits one to arise with parametric efficiency and the well-determined $K$-matrix pole parameter might suggest the presence of a $t$-matrix pole. Furthermore, parametrisations that do not permit poles were unable to successfully reproduce our finite-volume spectra. In Section~\ref{sec:swave_elastic_param_vars} we explore further forms of the $t$-matrix and interpret our results in terms of poles in Section~\ref{sec_poles}.

\subsubsection{$P$-wave}\label{sec:pwave_scat}

In each of the irreps with an $\ell=1$ contribution, we find an energy level at $a_t E_\cm \approx 0.35$ well below $D\pi$ threshold. The first source of possible inelasticity in $\ell=1$ comes from $D^\star\pi$ contributions, so we exclude extracted levels above $D^\star\pi$ threshold. We consider the ten energy levels below the $D^\star\pi$ (and also below the $D\pi$) threshold from irreps that have $\ell = 1$ as their lowest contributing partial wave, these are the black points in Fig. \ref{spec_P}. Assuming that contributions coming from
$\ell \geq 2$ are negligible, the $t$-matrix can be parametrised yielding a $P$-wave scattering amplitude around $a_t E_\cm \approx 0.35$. Using a $K$-matrix description that includes only a pole term, we obtain\\
\begin{center}
\begin{tabular}{rll}
$m_{1} \;\; =$ & $(0.35443 \pm 0.00021 \pm 0.00007) \cdot a_t^{-1}$ & \multirow{2}{*}{ $\begin{bmatrix} 1 & -0.73 \\ & 1\end{bmatrix}$ } \\
$g_{1} \;\; =$ & $1.58 \pm 0.32 \pm 0.02$  & \\[1.3ex]
&\multicolumn{2}{l}{ $\chi^2/ N_\mathrm{dof} = \frac{11.23}{10 - 2} = 1.40.$}  \\
\end{tabular}
\end{center}
\vspace{-1cm}
\begin{equation}\label{eq_P_fit}\end{equation}\\
All of the energy levels found at $a_t E_\cm \approx 0.35$ have a large overlap with $\bar{q}q$ operators subduced from those with $J=1$ in the infinite-volume continuum. Along with the well-determined pole parameter in Eq.~(\ref{eq_P_fit}), this may indicate the presence of a deeply bound vector state consistent with what was previously found in Ref.~\cite{Moir:2013ub} (which did not include multi-meson operators). We find no other levels in the elastic scattering region meaning that we can not constrain the $P$-wave scattering amplitude further without including irreps that have both $S$ and $P$-wave contributions.

\subsubsection{$S$ and $P$-waves}

\begin{figure}[t!]
\begin{center}
\includegraphics[width=0.8\textwidth]{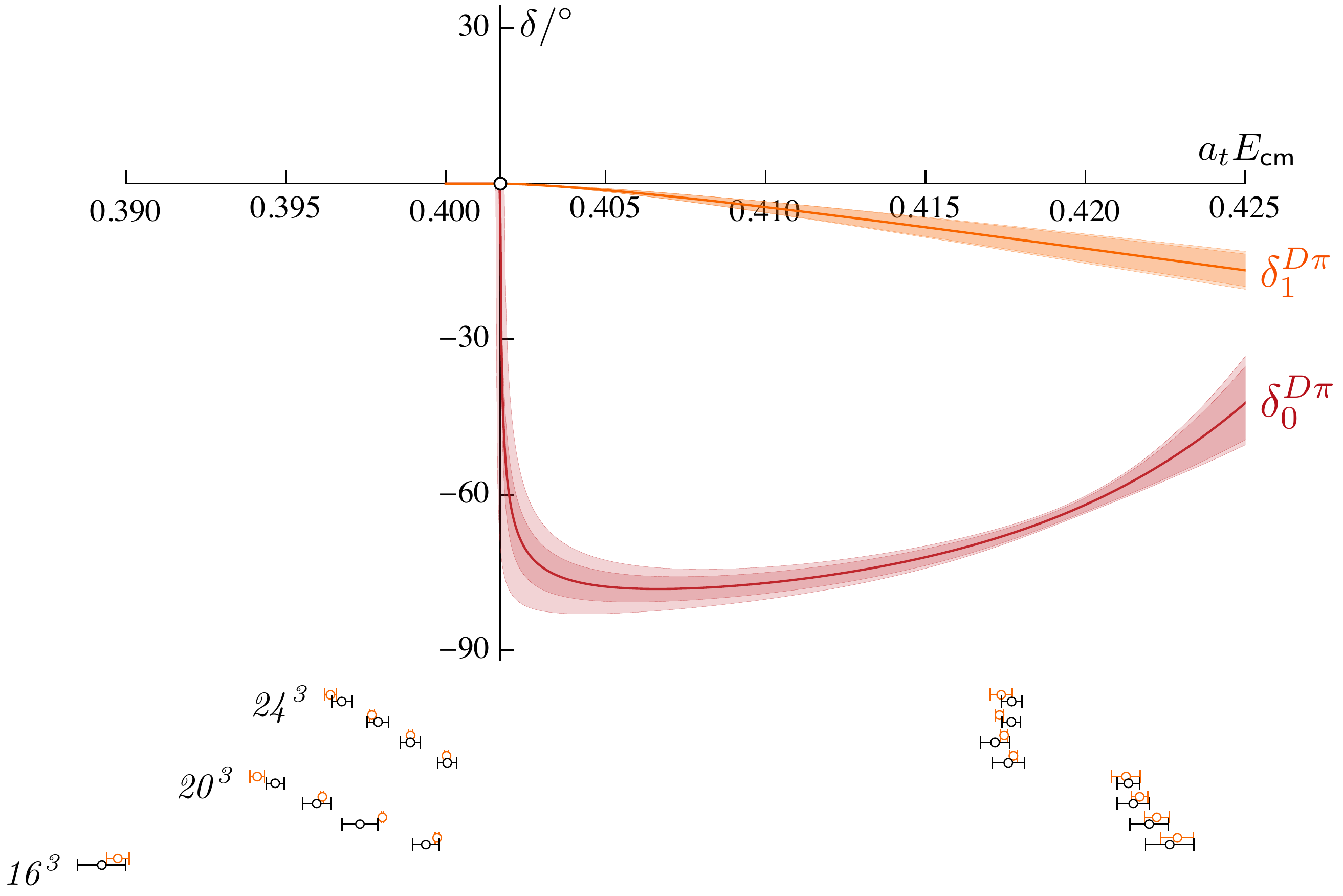}
\caption{The red (orange) band shows the $S$-wave ($P$-wave) phase shift obtained from the parametrisation in Eq.~(\ref{eq_fit_SP}). The inner band represents the one-sigma statistical uncertainty while the outer band shows the additional uncertainty coming from varying the scattered meson masses and the anisotropy within their statistical uncertainty. The open circle at $a_{t} E_{\cm} = 0.40171$ shows the location of the $D\pi$ threshold. The black points show the location of our finite-volume energy levels while the orange points show those corresponding to the parametrisation in Eq.~(\ref{eq_fit_SP}).}
\label{fig_elastic_S_single}
\end{center}
\end{figure}

We now determine $S$ and $P$-wave scattering amplitudes simultaneously. To do this, we make use of our finite-volume energy levels below the $D\eta$ threshold in the $[000]A_{1}^{+}$ irrep, and below the $D^{\star}\pi$ threshold in the $[000]T_{1}^{-}$, $[\vec{P} \neq 0]A_{1}$, $[001]E_{2}$, $[011]B_{1}$, $[011]B_{2}$ and $[111]E_{2}$ irreps; there are 33 in total.  As justified in Section~\ref{sec:Dwave_param_vars}, we neglect $\ell \geq 2$ contributions.  Separately for each of the $\ell = 0$ and $\ell=1$ parts of the $t$-matrix, we use a $K$-matrix containing a pole term and a constant. This leads to
\begin{center}
\begin{tabular}{rll}
$m \;\;=$                         & $(0.393 \pm 0.002 \pm 0.002) \cdot a_t^{-1}$   & 
\multirow{6}{*}{ $\begin{bmatrix} 1.00 & -0.16 &   0.67  &  0.18  &  0.14   &  0.47 \\ 
                                    &  1.00    &   0.61  & -0.14  &  0.18   &  0.21 \\
                                    &          &   1.00  &  0.02  &  0.23   &  0.49 \\   
                                    &          &         &  1.00  & -0.69   & -0.35 \\
                                    &          &         &        &  1.00   &  0.61 \\
                                    &          &         &        &         &  1.00    \end{bmatrix}$ } \\
$g \;\;=$                 & $(0.60 \pm 0.02 \pm 0.05) \cdot a_t^{-1}$   & \\
$\gamma \;\;= $           & $\,\,\,\;10.1 \pm 1.1 \pm 1.0 $   & \\
$m_1 \;\;=$                 & $(0.35444 \pm 0.00014 \pm 0.00004) \cdot a_t^{-1}$   & \\
$g_1 \;\;=$                 & $1.45 \pm 0.31 \pm 0.04 $    & \\
$\gamma_1 \;\;= $           & $(-104 \pm 44 \pm 6) \cdot a_t^{2} $   & \\[1.3ex]
&\multicolumn{2}{l}{ $\chi^2/ N_\mathrm{dof} = \frac{44.2}{33-6} = 1.64 $\,,}
\end{tabular}
\end{center}
\vspace{-0.8cm}
\begin{equation} \label{eq_fit_SP}\end{equation}\\
where the parameters with a subscript $1$ denote the $P$-wave.

In Fig.~\ref{fig_elastic_S_single}, we show the phase shifts, $\delta^{D\pi}_{0}$ and $\delta^{D\pi}_{1}$, along with the finite-volume energy levels (black points) and those given by the parametrisation in Eq.~(\ref{eq_fit_SP}) (orange). Although the points clustered around $a_tE_\cm \approx 0.35$ are not shown, they are included in our fit. Just above the $D\pi$ threshold the $S$-wave phase shift shows a rapid variation, a feature indicative of a nearby pole. On the other hand, the $P$-wave phase shift varies slowly throughout the energy range shown.

\begin{figure}[t!]
\begin{center}
\includegraphics[width=0.66\textwidth]{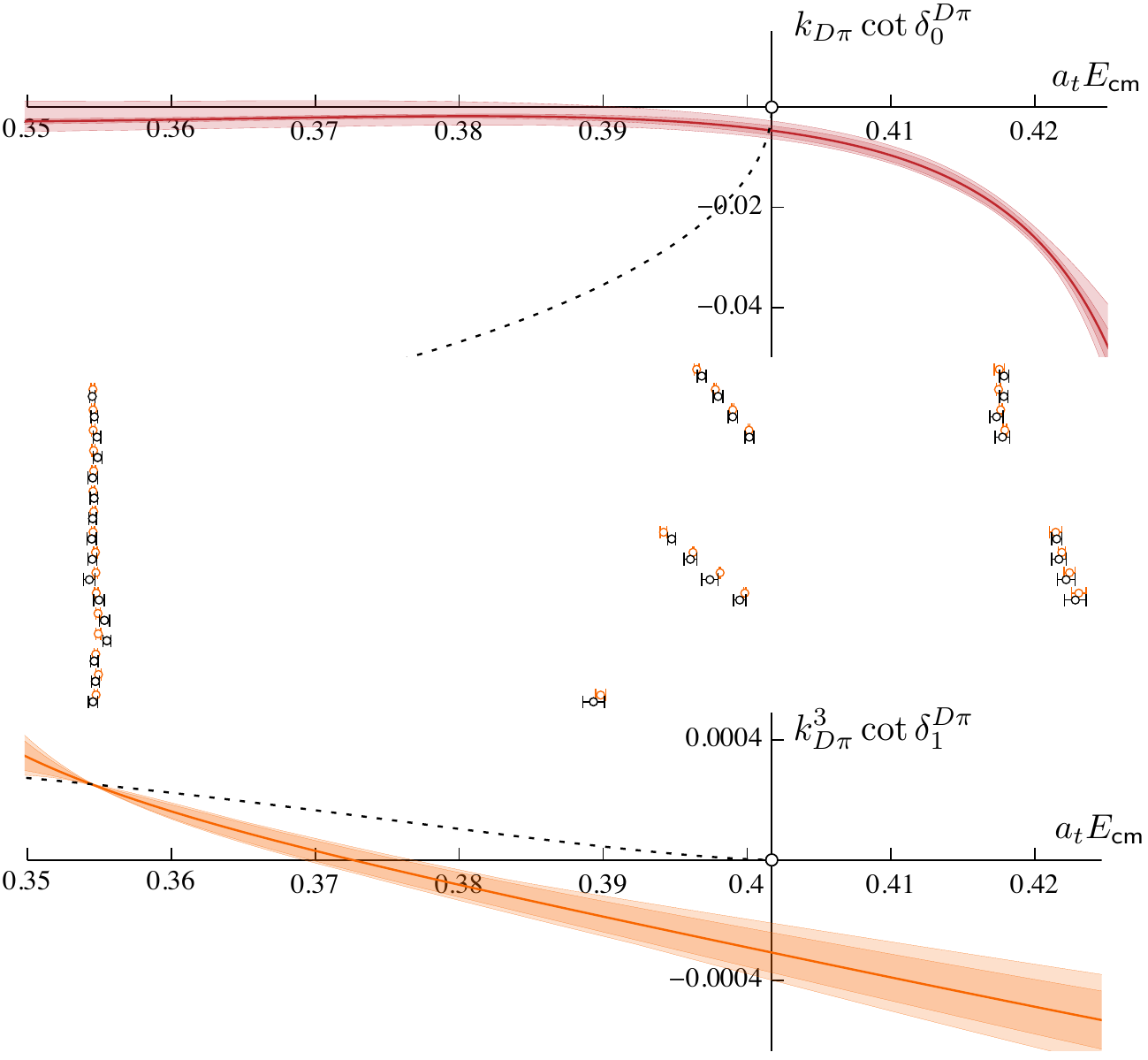}
\caption{As Fig.~\ref{fig_elastic_S_single} but for the quantities $k_{D\pi} \cot \delta^{D\pi}_{0}$ and $k^{3}_{D\pi} \cot \delta^{D\pi}_{1}$. The dotted curves show $ik_\pD^{2\ell+1}$. }
\label{fig_elastic_kcotdelta}
\end{center}
\end{figure}

In the upper (lower) panel of Fig.~\ref{fig_elastic_kcotdelta}, we show the quantity $k_{D\pi} \cot \delta^{D\pi}_{0}$ ($k^{3}_{D\pi} \cot \delta^{D\pi}_{1}$) determined from the parametrisation in Eq.~(\ref{eq_fit_SP}). The dotted curves correspond to the quantity $ik_{D\pi}^{2\ell+1}$, which should intersect the bands at the location of a subthreshold $t$-matrix pole on the physical sheet. The intersection with $k_{D\pi} \cot \delta^{D\pi}_{0}$ provides evidence that the possible pole near the $D\pi$ threshold actually lies just below it, while the intersection with $k^{3}_{D\pi} \cot \delta^{D\pi}_{1}$ suggests a bound state around $a_tE_\cm \approx 0.35$.  We defer further discussion to Section~\ref{sec_poles} where we investigate the singularity content of these scattering amplitudes.

\subsubsection{Parametrisation Variation}\label{sec:swave_elastic_param_vars}

\begin{figure}[t!]
\begin{center}
\includegraphics[width=0.95\textwidth]{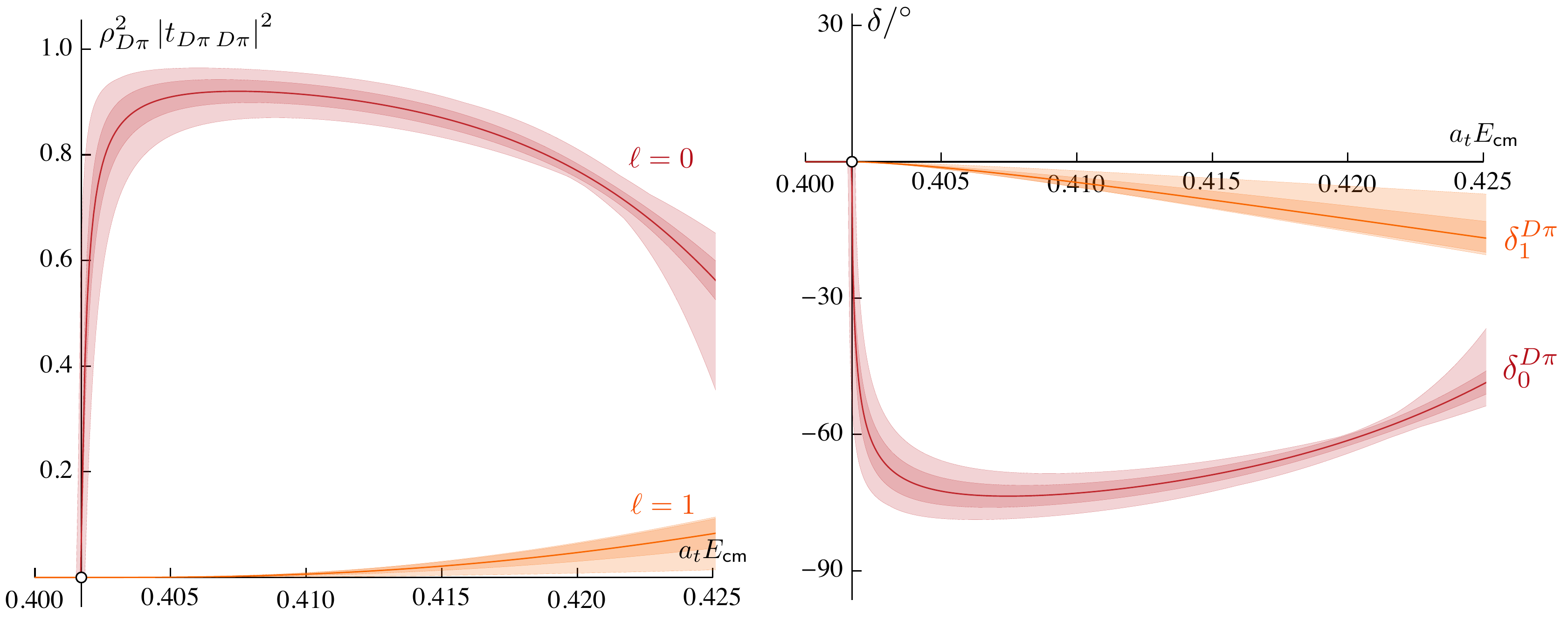}
\caption{The quantity $\rho_{D\pi}^{2} |t_{D\pi D\pi}|^2$ (left) and the phase shift $\delta^{D\pi}_{\ell}$ (right). The red and orange bands correspond to $S$ and $P$-wave respectively -- they encompass the various parametrisations with $\chi^2/N_\mathrm{dof} < 1.9$ in Table~\ref{tab_elastic_variations} as well as the uncertainty from the anisotropy and the scattered meson masses.} 
\label{fig_elastic_S_kcot_amp_par_var}
\end{center}
\end{figure}

To assess the extent to which the scattering amplitudes depend upon the choice of parametrisation, we repeat the procedure above for a variety of forms of both the $S$ and $P$-wave parts of the $t$-matrix. A selection of these forms is summarised in Table~\ref{tab_elastic_variations} in Appendix~\ref{app:parametrisation}, where we also show the $\chi^2/N_\mathrm{dof}$ obtained when minimising our $\chi^2$ function using the same 33 energy levels as above. Several other parametrisations have also been tested, such as higher orders of the forms shown or the inverse polynomial $K$-matrix used in Ref.~\cite{Wilson:2014cna}; these are found to be highly correlated or contained parameters consistent with
zero so we do not show them.

In Fig.~\ref{fig_elastic_S_kcot_amp_par_var}, we show the quantity $\rho_{D\pi}^{2} |t_{D\pi D\pi}|^2$ (left), which is proportional to the $D\pi \to D\pi$ cross-section, and the phase shift $\delta^{D\pi}_{\ell}$ (right). The size of the bands encompasses the variation and uncertainty coming from all parametrisations with $\chi^2/N_\mathrm{dof} < 1.9$. From a comparison of the phase shifts with Fig.~\ref{fig_elastic_S_single}, it is clear that the parametrisation dependence is almost negligible. In the quantity $\rho_{D\pi}^{2} |t_{D\pi D\pi}|^2$, we observe further evidence of an $S$-wave pole near the $D\pi$ threshold, as the large ``peak" almost saturates the unitarity bound, which is unity in our normalisation.

\subsection{Coupled-Channel Scattering in $S$-wave}\label{sec:coupled_SP}

\begin{figure}[t!]
\begin{center}
\includegraphics[width=0.6\textwidth]{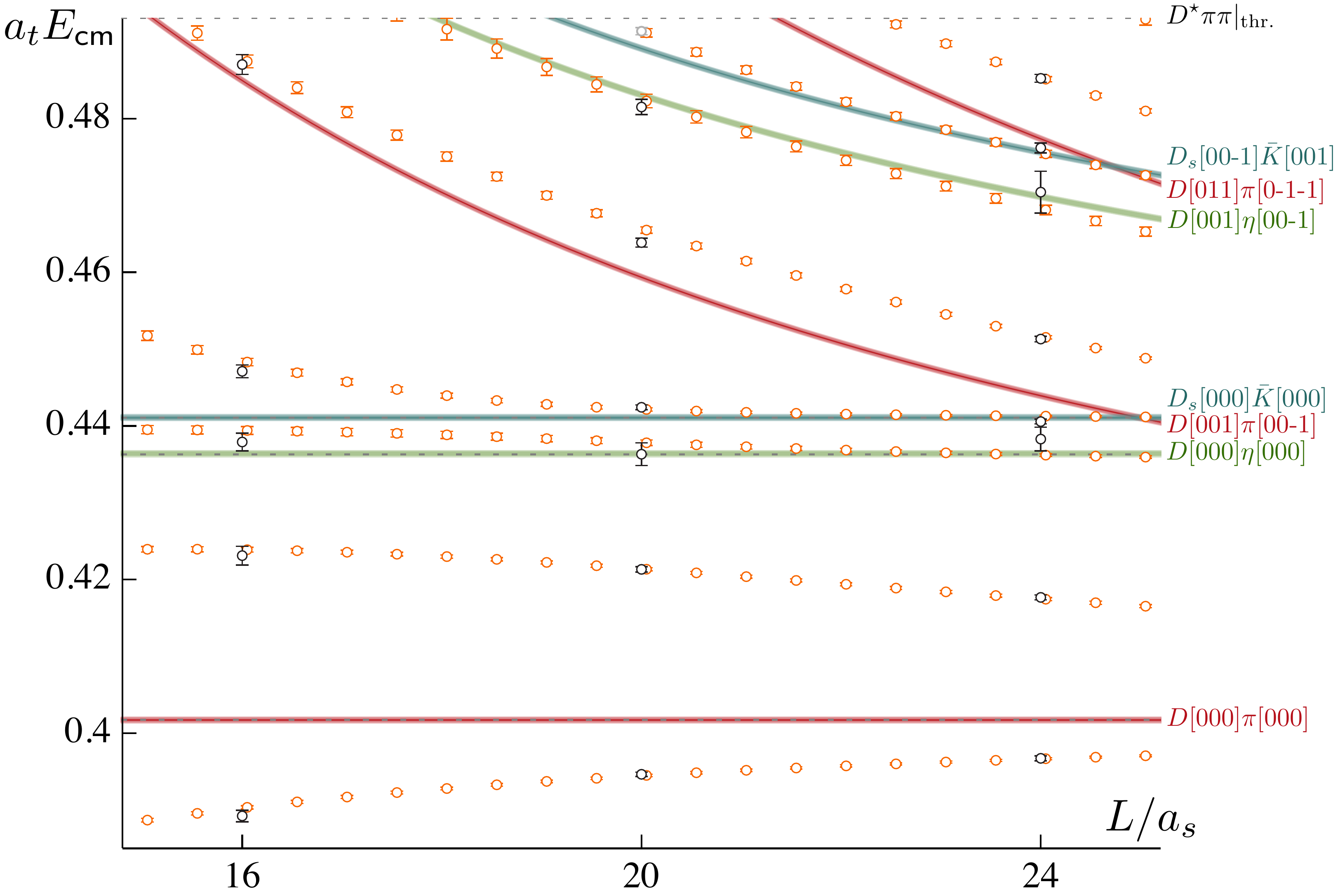}
\caption{A comparison between the finite-volume spectrum in the $[000]A^{+}_{1}$ irrep (black points) and the spectrum coming from the parametrisation in Eq.~(\ref{fit_minimum_3x3_S}) (orange points). The red, green and blue curves show the location of non-interacting $D\pi$, $D\eta$ and $D_{s}\bar{K}$ energies respectively, while the grey dotted line shows the threshold for the lowest channel for which we have not included operators in the variational procedure, namely $D^{\star}\pi\pi$.}
\label{fig_spectra_fitted_A1p}
\end{center}
\end{figure}

We now go beyond the the elastic $D\pi$ energy region and consider the case of coupled-channel $D\pi$, $D\eta$ and $D_{s}\bar{K}$ scattering in $S$-wave in combination with elastic $D\pi$ scattering in $P$-wave; the $P$-wave constraint comes only from energy levels below the coupled-channel region. The energy levels in the coupled-channel region are all from $[000]A_1^+$ where the first contamination comes from $\ell=4$ and is expected to be highly suppressed.  The irreps with $\vec{P} \neq \vec{0}$ can have contributions from $\ell=2$ but these are later shown to be negligible in the elastic region.
Using the 47 energy levels coloured black in Figs.~\ref{spec_A1} and \ref{spec_P}, we parametrise the $t$-matrix by a coupled-channel $K$-matrix in $S$-wave and an elastic $K$-matrix in $P$-wave, giving
\begin{center}\small
\begin{tabular}{rll}
$m =$                       & $\!\!\!\! (0.40161 \pm 0.00006 \pm 0.00007) \cdot a_t^{-1}$   & 
\multirow{8}{*}{ $\!\!\!\!
                  \begin{bmatrix}     1 & \sm0.15 & \sm 0.02 &  0.04   & \sm 0.07 & \sm 0.02 &  0.03    & \sm 0.28   \\
                                      	&  1      & \sm 0.94 & \sm 0.25 &  0.75   & \sm 0.80 &  0.79    & \sm 0.25   \\
                                      	&         &  1      &  0.44   & \sm 0.74  &  0.87    & \sm 0.89 &     0.26   \\
                                      	&         &         &  1      & \sm 0.41  &  0.55    & \sm 0.63 & \sm 0.01   \\
                                      	&         &         &         &  1        & \sm 0.94 &  0.78    & \sm 0.51   \\
                                      	&         &         &         &           &  1       & \sm 0.93 &     0.47   \\
                                        &         &         &         &           &          &  1       & \sm 0.33   \\
                                        &         &         &         &           &          &          &     1     \\
                                       \end{bmatrix}$ } \\
$g_{\pD}                = $ & $\!\!\!\! (\,\,\,\;  0.62 \pm 0.04 \pm 0.05) \cdot a_t^{-1}$   & \\
$g_{\eD}                = $ & $\!\!\!\! (         -0.52 \pm 0.07 \pm 0.10) \cdot a_t^{-1}$   & \\
$g_{\KDs}               = $ & $\!\!\!\! (\,\,\,\;  0.23 \pm 0.03 \pm 0.04) \cdot a_t^{-1}$   & \\
$\gamma_{\pD,  \,\pD }  = $ & $\!\!\!\!\,\,\,\;  2.3 \pm 0.8 \pm 1.0 $   & \\
$\gamma_{\pD,  \,\eD }  = $ & $\!\!\!\!         -1.6 \pm 0.9 \pm 1.2 $   & \\
$\gamma_{\eD,  \,\eD }  = $ & $\!\!\!\!\,\,\,\;  2.7 \pm 1.0 \pm 1.5 $   & \\
$\gamma_{\KDs,  \,\KDs}=  $ & $\!\!\!\!         -0.3 \pm 0.2 \pm 0.2 $   & \\
$m_1 =$                 & $\!\!\!\! (0.35459 \pm 0.00016 \pm 0.00004) \cdot a_t^{-1}$    & \hspace{-0.32cm} \multirow{3}{*}{ $\begin{bmatrix} 1.00 & -0.75    &  -0.42  \\ 
                                                                                                                             &  1.00    &   0.59  \\
                                                                                                                             &          &   1.00  \end{bmatrix}$ } \\
$g_1 \;\;=$                 & $1.30 \pm 0.36 \pm 0.07 $   & \\
$\gamma_1 \;\;= $           & $(-94 \pm 35 \pm 4) \cdot a_t^{2} $   & \\[1.5ex]
&\multicolumn{2}{l}{ $\chi^2/ N_\mathrm{dof} = \frac{61.6}{47-11} = 1.71 $ \, .}
\end{tabular}
\vspace{-0.6cm}
\begin{equation}\label{fit_minimum_3x3_S}\end{equation}
\end{center}\normalsize
Note that the $P$-wave parameters are consistent with what is obtained in Eq.~(\ref{eq_fit_SP}); as one might expect, the inclusion of the coupled-channel region in $S$-wave appears to have a negligible effect on the $P$-wave amplitudes. In Fig.~\ref{fig_spectra_fitted_A1p} we show a comparison between our finite-volume spectrum in the $[000]A_{1}^{+}$ irrep (black points) and the spectrum coming from the parameters in Eq.~(\ref{fit_minimum_3x3_S}) (orange points). 

\begin{figure}[t!]
\begin{center}
\includegraphics[width=0.49\textwidth]{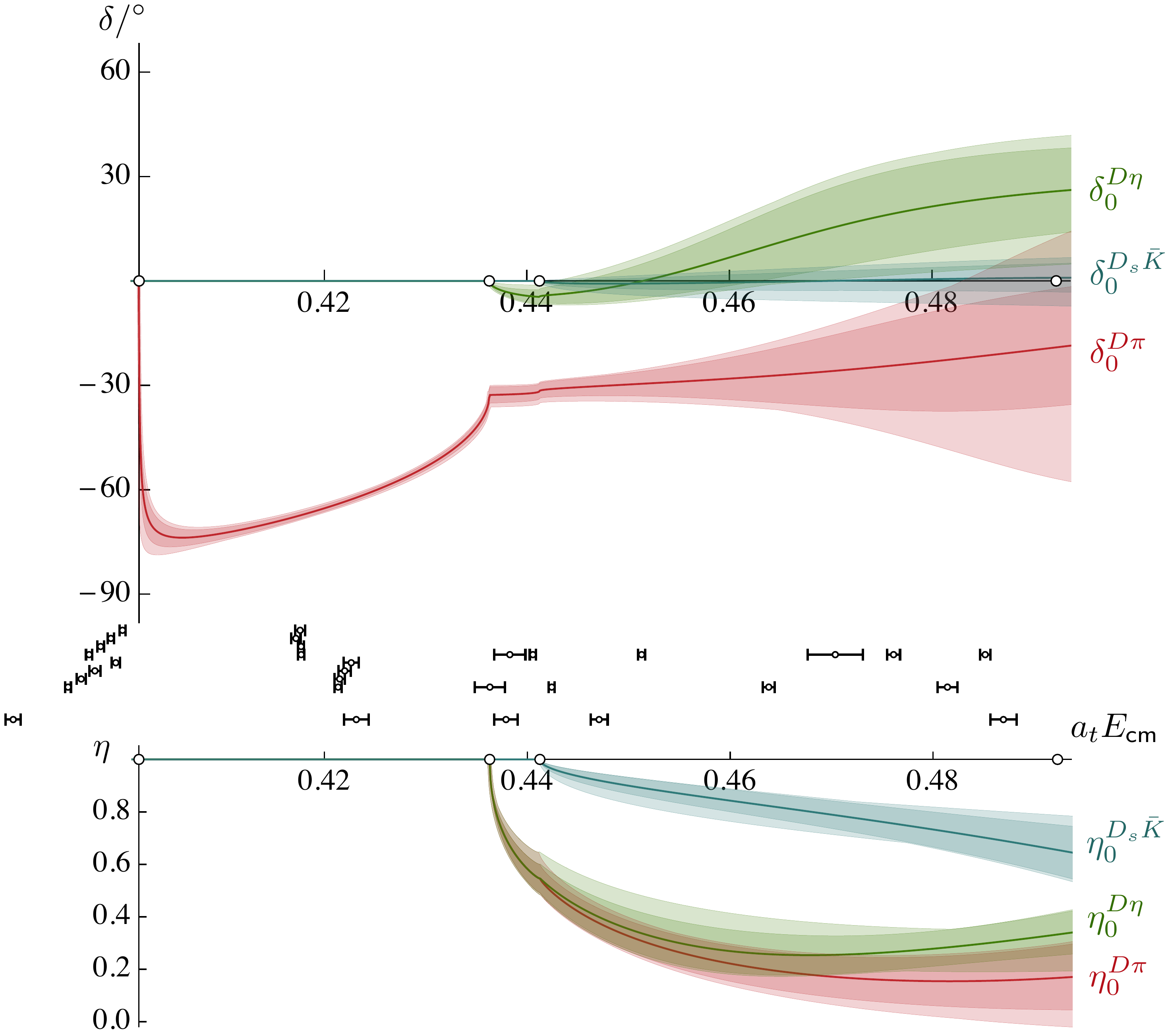}
\includegraphics[width=0.49\textwidth]{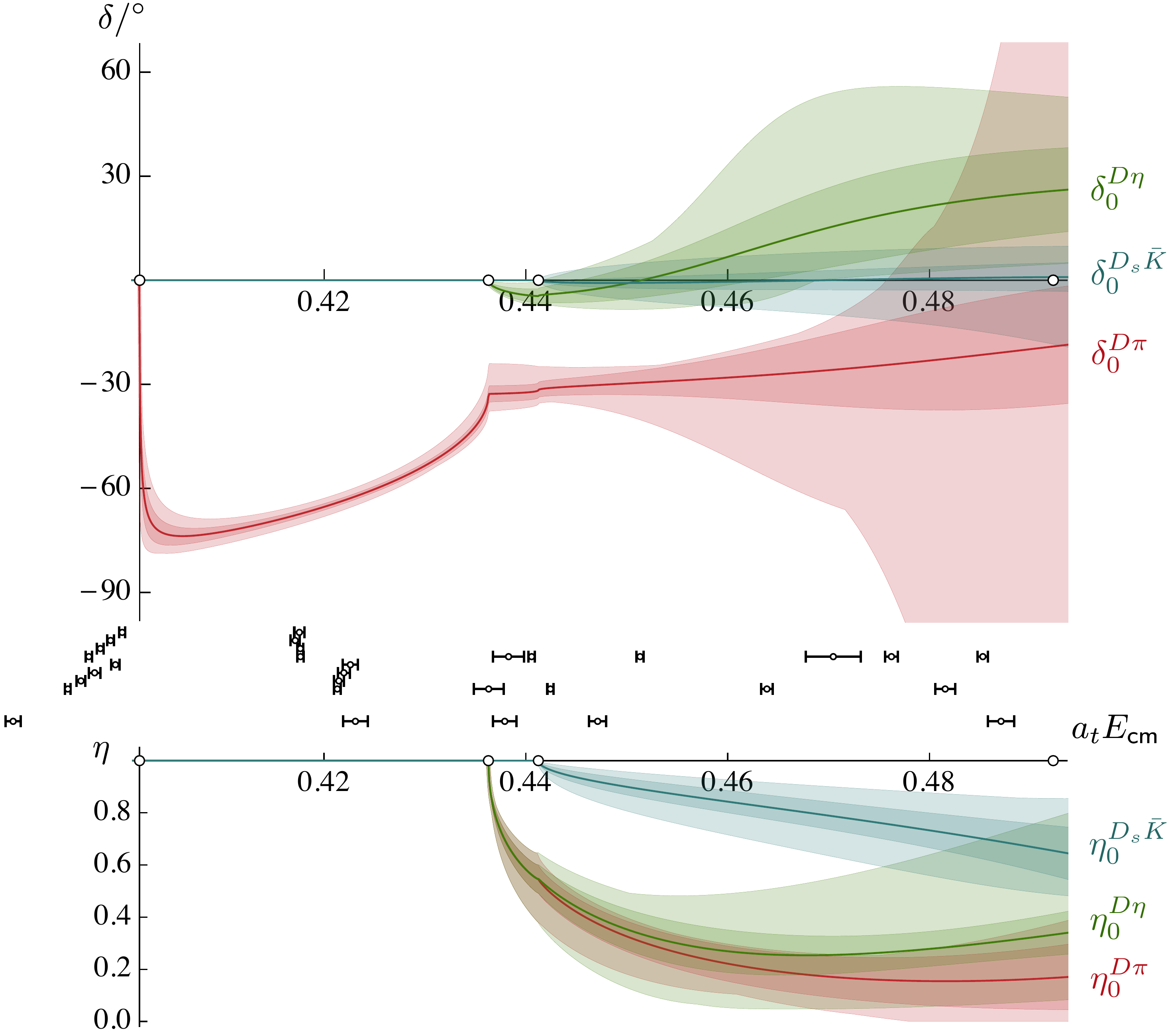}
\caption{The upper (lower) left panel shows the $S$-wave phase shifts (inelasticities) for the $D\pi$ (red), $D\eta$ (green) and $D_{s}\bar{K}$ (blue) channels, determined from the parametrisation in Eq.~(\ref{fit_minimum_3x3_S}). The upper (lower) right panel shows the same phase shifts (inelasticities) where the size of the bands incorporates all parametrisations shown in Table~\ref{tab_coupled_variations} with $\chi^2/N_\mathrm{dof} < 1.9$. The black points between the upper and lower panels show the location of the finite-volume energy levels used to constrain the parametrisations.}
\label{fig_coupled_phase}
\end{center}
\end{figure}

In the upper left panel of Fig.~\ref{fig_coupled_phase}, we show the $S$-wave phase shifts $\delta^{D\pi}_{0}$ (red), $\delta^{D\eta}_{0}$ (green), and $\delta^{D_{s}\bar{K}}_{0}$ (blue) corresponding to the parametrisation in Eq~(\ref{fit_minimum_3x3_S}). By comparing to the elastic case in the right panel of Fig.~\ref{fig_elastic_S_kcot_amp_par_var}, we see that our results for the elastic $D\pi$ region are largely unaffected when we allow for the $D\eta$ and the $D_{s}\bar{K}$ channels. However, at the opening of the $D\eta$ threshold we do observe a noticeable ``kink'' in the $D\pi$ phase shift suggesting a non-zero coupling between the two channels. We see a much smaller effect at the opening of the $D_{s}\bar{K}$ threshold. The non-zero coupling between channels is further demonstrated in the lower left panel of Fig.~\ref{fig_coupled_phase}, which shows a clear deviation of the inelasticities from unity. 

The upper (lower) left panel of Fig.~\ref{fig_coupled_amp} shows $\rho_i\rho_{j} |t_{ij}|^2$ for $i = j$ ($i \neq j$) determined from the parametrisation in Eq~(\ref{fit_minimum_3x3_S}); this quantity is proportional to the cross section for scattering of channel $i \to i$ ($i \to j$). We see that just above $D\pi$ threshold, as in the elastic case, the unitarity bound is almost saturated for $D\pi \to D\pi$.

\subsubsection{Parametrisation Variation}\label{sec:coupled_scat_swave}

We now assess the extent to which our results depend upon our choice of parametrisation of the $t$-matrix. Table~\ref{tab_coupled_variations} in Appendix~\ref{app:op_tables} shows a selection of parametrisations of the $t$-matrix we considered with the $\chi^2 / N_\mathrm{dof}$ obtained in each case. Note that, we have also attempted several other parametrisations, such as $K_{ij}^{-1}=\sum c_{ij}^{(k)}s^k$ and those with higher order terms of the forms shown. However, these were found to either have insufficient freedom to describe our finite-volume spectra or to give highly correlated parameters.

In the upper (lower) right panel of Fig.~\ref{fig_coupled_phase}, we show the $D\pi$, $D\eta$ and $D_{s}\bar{K}$ $S$-wave phase shifts (inelasticities) where the size of the bands include the one-sigma statistical uncertainty coming from all parametrisations with a $\chi^2/N_\mathrm{dof} < 1.9$ as well as the statistical uncertainty coming from the scattered meson masses and the anisotropy. It appears that there is almost no parametrisation dependence up to around $a_{t}E_{\cm} \approx 0.46$, with all of the features described in the previous section remaining intact. Above this, we do not have many energy levels to constrain the scattering amplitude and hence we see a dramatic reduction on the constraint we can place on the phases\footnote{The complete loss of constraint in the $D\pi$ phase shift above $a_{t}E_{\cm} \approx 0.47$ is due to inelasticity being consistent with zero [see Eq.~(\ref{eq_phases_3x3})].}. 

\begin{figure}[t!]
\begin{center}
\includegraphics[width=0.49\textwidth]{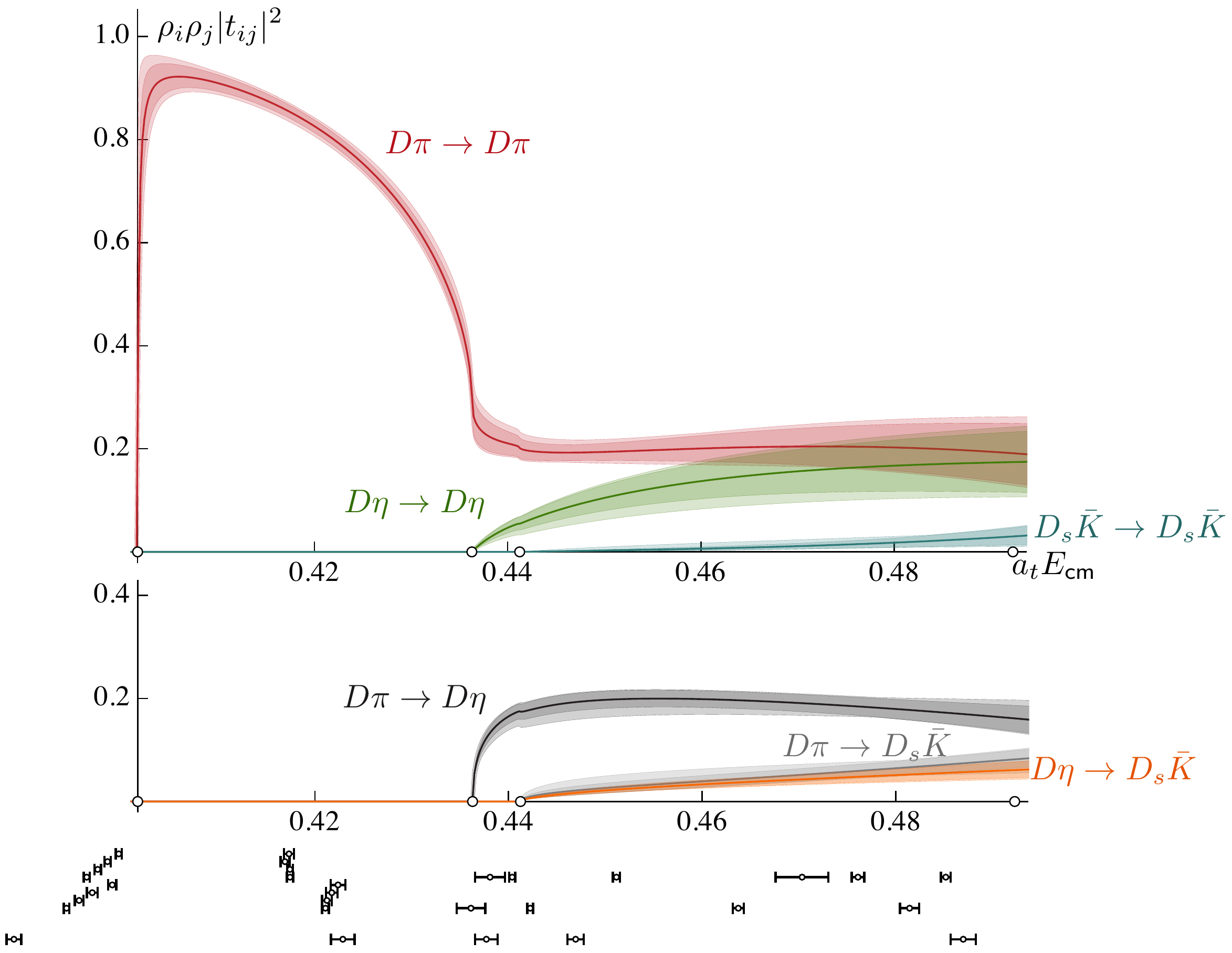}
\includegraphics[width=0.49\textwidth]{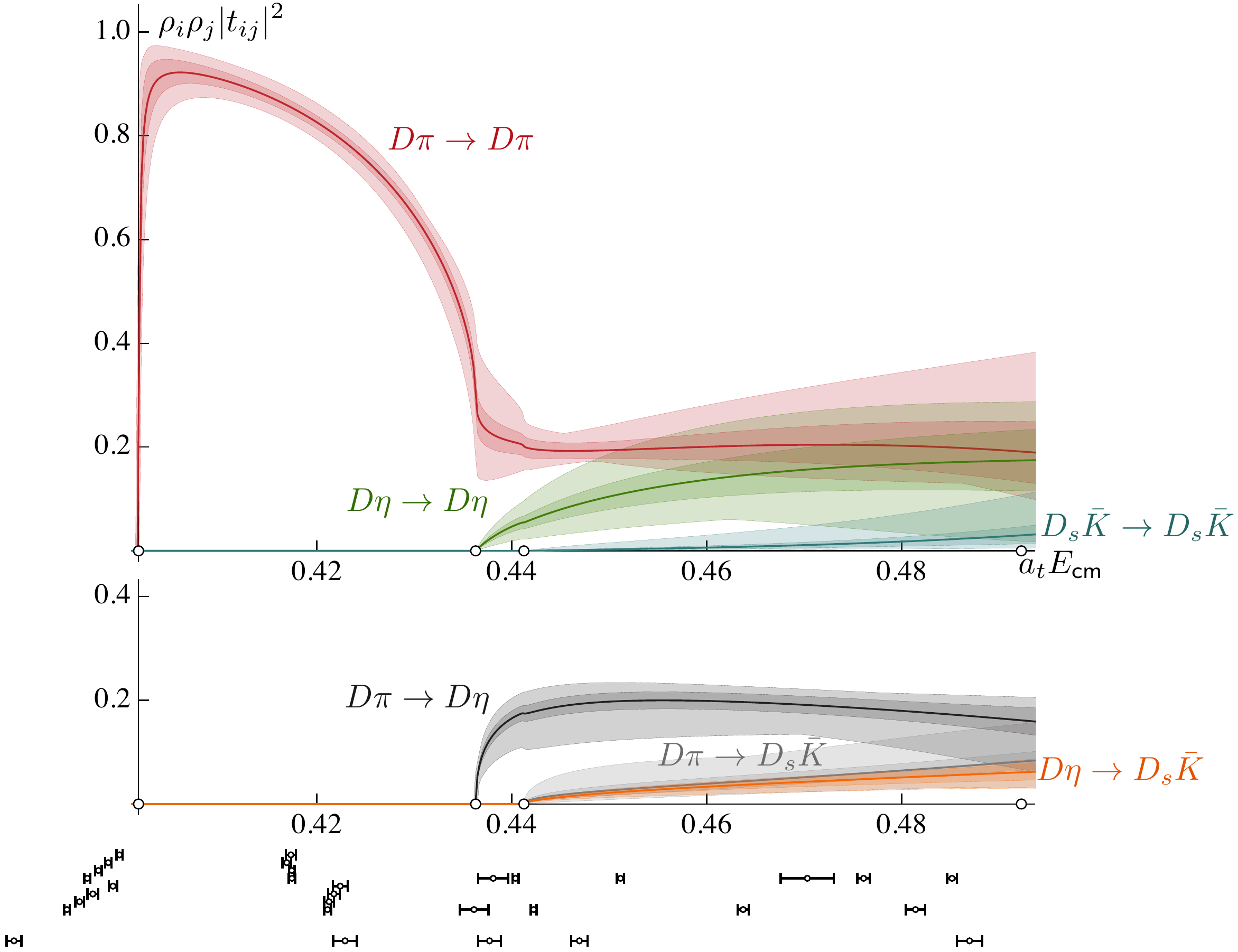}
\caption{The left panels show the quantity $\rho_i\rho_j |t_{ij}|^2$ determined using the parametrisation in Eq.~(\ref{fit_minimum_3x3_S}); the size of the bands include the one-sigma statistical uncertainty and the statistical uncertainty coming from the scattered meson masses and the anisotropy. The right panels show the same quantity where the bands now encompass all the parametrisations in Table~\ref{tab_coupled_variations} with $\chi^2/N_\mathrm{dof} < 1.9$. The black points show the location of the finite-volume energy levels used to constrain the parametrisations.}
\label{fig_coupled_amp}
\end{center}
\end{figure}

In the right panels of Fig.~\ref{fig_coupled_amp}, we show the quantity $\rho_i\rho_j |t_{ij}|^2$, where the size of the bands include the parametrisations from Table \ref{tab_coupled_variations} with $\chi^2/N_\mathrm{dof} < 1.9$. There appears to be very little parametrisation dependence in this quantity with all features, including the large ``peak" in the $D\pi \rightarrow D\pi$ channel just above the $D\pi$ threshold, remaining intact.

\subsection{Coupled-Channel Scattering in $D$-wave}\label{sec:Dwave_param_vars}

We now turn our attention to coupled-channel scattering in $D$-wave. The spectra shown in Fig.~\ref{spec_D}, namely $[000]E^+$, $[000]T_2^+$, $[001]B_1$ and $[001]B_2$, have $\ell = 2$ as the lowest contributing partial wave. Although we have included operators for many of the open channels within the energy region shown, there are some for which we have not included operators, notably the relatively low-lying $D^{\star}\pi$ channel. Nevertheless, $D$-wave channels are known to open slowly since they are suppressed in proportion to $k^{2\ell+1}$, suggesting that it may still be possible to apply the L\"{u}scher formalism; Ref.~\cite{Giudice:2012tg} showed that, at least in $S$-wave, the breakdown of the formalism above an inelastic threshold results in values of the phase shift clearly incompatible with those below the threshold. We have thoroughly checked for the presence of such effects but do not find any evidence for them. Encouraged by this, we cautiously proceed with our scattering analysis. 

\begin{figure}
\begin{center}
\includegraphics[width=0.95\textwidth]{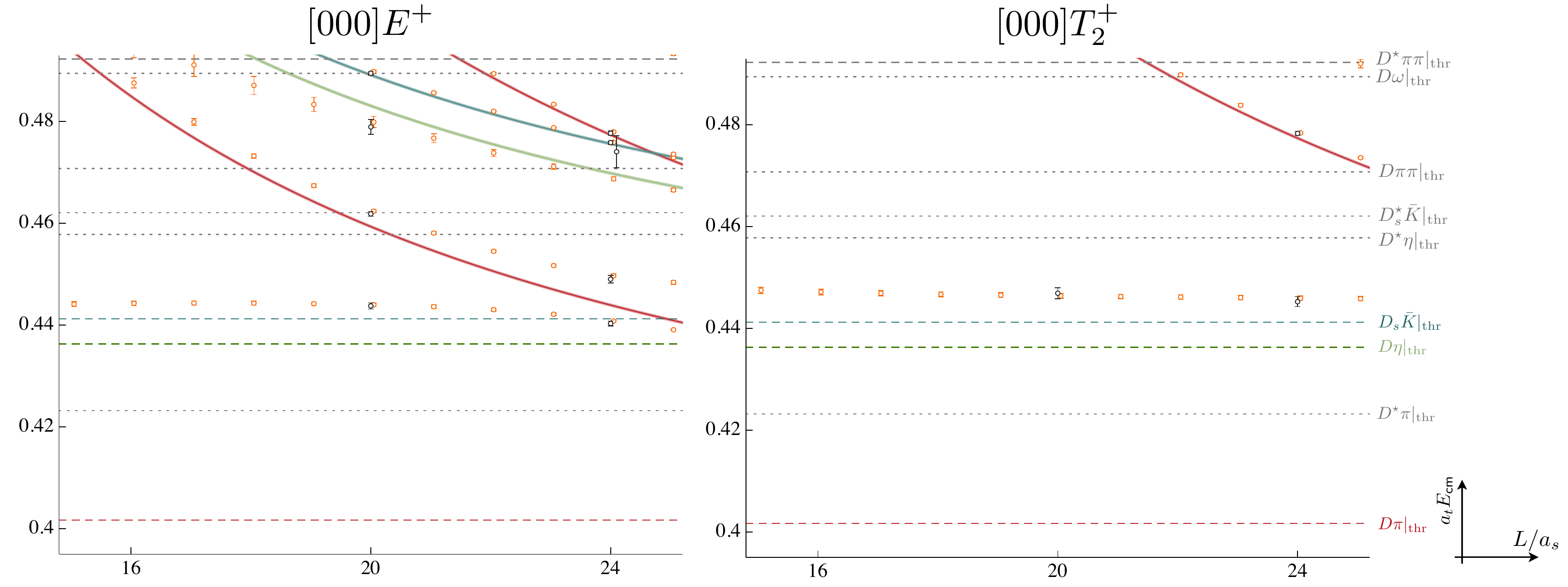}
\caption{As Fig.~\ref{fig_spectra_fitted_A1p} but for the $[000]E^{+}$ and $[000]T^{+}_{2}$ irreps and with the orange points coming from the parametrisation in Eq.~(\ref{fit_minimum_3x3_D}).}
\label{fig_Dwave_par_spec}
\end{center}
\end{figure}

Utilising the 28 energy levels up to around the $D^{\star}\pi\pi$ threshold in Fig.~\ref{spec_D}, we parametrise the $D$-wave part of the $t$-matrix using a three-channel $K$-matrix form. We find a reasonable fit with
\begin{center}
\small
\begin{tabular}{rll}
$m = \!\!$                       & $\!\! (0.44590 \pm 0.00048 \pm 0.00006) \cdot a_t^{-1}$   & \hspace{-0.2cm}
\multirow{7}{*}{ $\begin{bmatrix}     1 &  0.35 & -0.04 &  0.09 & -0.11 & -0.31 & -0.30 \\
                                      	&  1    & -0.07 &  0.10 &  0.54 & -0.20 & -0.10 \\
                                      	&       &  1    & -0.22 &  0.01 &  0.46 &  0.16 \\
                                      	&       &       & 1     & -0.01 & -0.16 & -0.89 \\
                                      	&       &       &       &  1    &  0.15 &  0.19 \\
                                      	&       &       &       &       &  1    &  0.23 \\ 
                                        &       &       &       &       &       &  1
                                       \end{bmatrix}$ } \\
$g_{\pD}                = \!\!$  & $\!\! (\,\,\,\; 1.766 \pm 0.049 \pm 0.009) \cdot a_t$   & \\
$g_{\eD}                = \!\!$  & $\!\! (        -0.60 \pm 0.91 \pm 0.04) \cdot a_t$   & \\
$g_{\KDs}               = \!\!$  & $\!\! (        -0.80 \pm 0.98 \pm 0.07) \cdot a_t$   & \\
$\gamma_{\pD,  \,\pD }  = \!\!$  & $\!\! (\,\,\,\; 40 \pm 16 \pm 8   ) \cdot a_t^4$   & \\
$\gamma_{\eD,  \,\eD}   = \!\!$  & $\!\! ( \,\,    294 \pm 83 \pm 22 ) \cdot a_t^4$   & \\
$\gamma_{\KDs, \,\KDs}  = \!\!$  & $\!\! (         -20 \pm 46 \pm 9  ) \cdot a_t^4$   & \\[3ex]
&\multicolumn{2}{l}{ $\chi^2/ N_\mathrm{dof} = \frac{25.4}{28-7} = 1.21 $\,.}
\end{tabular}
\end{center}
\vspace{-0.7cm}
\begin{equation} \label{fit_minimum_3x3_D}\end{equation}
In Fig.~\ref{fig_Dwave_par_spec}, we compare the finite-volume spectra in the $[000]E^+$ and $[000]T^{+}_{2}$ irreps to those determined from the parametrisation in Eq.~(\ref{fit_minimum_3x3_D}).

\begin{figure}
\begin{center}
\includegraphics[width=1.0\textwidth]{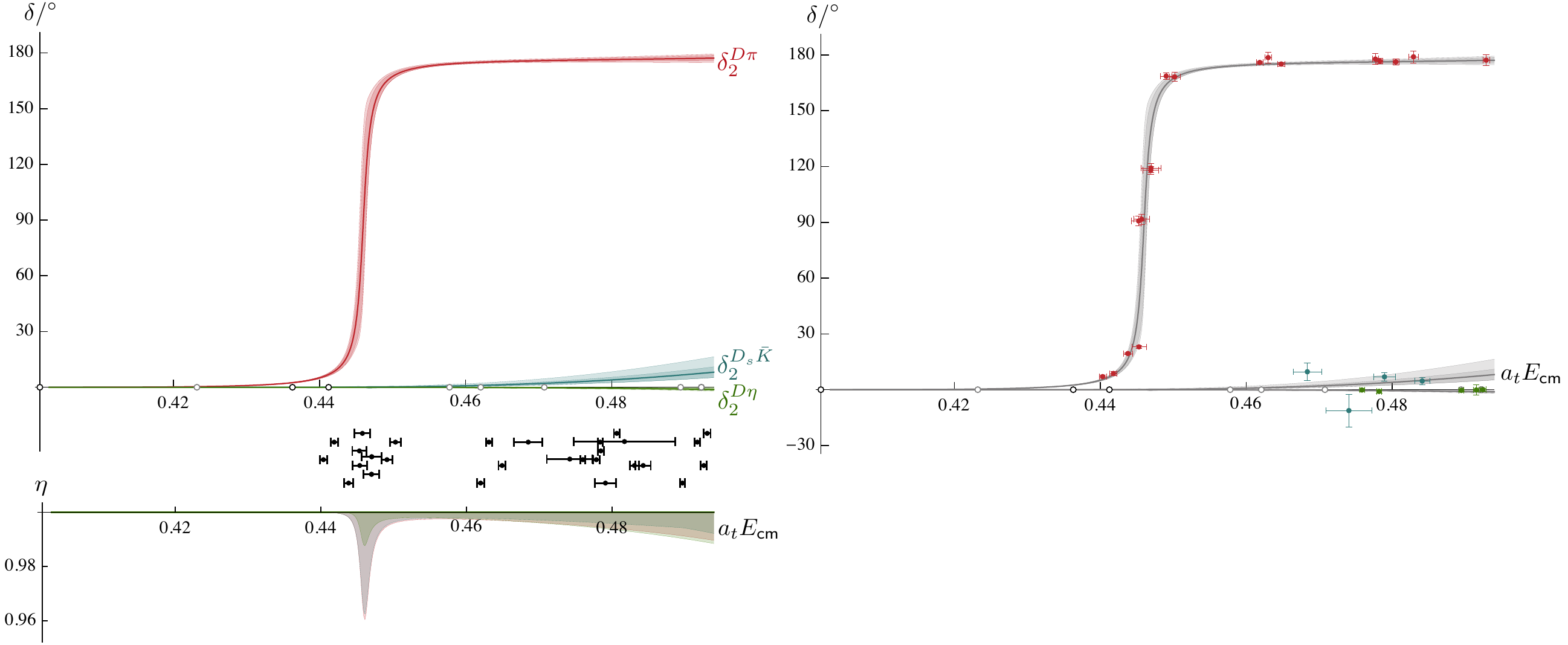}
\caption{The upper (lower) left panel shows the $D$-wave phase shifts (inelasticities), $\delta^{i}_{2}$ ($\eta^{i}_{2}$), where the size of the bands include all of the parametrisation variations listed in Table~\ref{tab_D_wave_par_vars}. The right panel shows the same phase-shift bands superimposed with points determined for each energy level as described in the text.}
\label{fig_D_phases}
\end{center}
\end{figure}

As before, we assess the extent to which our results depend upon a given parametrisation; Table~\ref{tab_D_wave_par_vars} in Appendix~\ref{app:op_tables} shows a selection of parametrisations of the $K$-matrix used to determine the $D$-wave scattering amplitude. The upper left panel of Fig.~\ref{fig_D_phases} shows the phase shift, $\delta^{D\pi}_{2}$, where the size of the band incorporates all of the parametrisations shown in Table~\ref{tab_D_wave_par_vars}. 
We observe that the phase shifts are small and well determined in the elastic region, justifying our neglect of $D$-wave contributions when determining the $S$ and $P$-wave amplitudes above.
The lower left panel shows the corresponding inelasticities, where a clear decoupling of the channels is observed. This enables a one-to-one correspondence between the finite-volume energy levels and the phases. In the right panel of Fig.~\ref{fig_D_phases}, we show the phase shift points determined for each energy level superimposed onto the phase shifts from the upper left panel\footnote{We have assigned a given finite-volume energy level to a single channel based on its dominant operator overlaps.}. The agreement between the two approaches further indicates the lack of coupling of the $D\eta$ and $D_{s}\bar{K}$ channels to the resonance.  

The narrowness of the phase shift and the apparent decoupling of the channels suggest that a Breit-Wigner parametrisation may also be capable of describing the resonance. By selecting only those levels identified as belonging to the $D\pi$ channel, we obtain a reasonable description with the parameters,
\begin{center}
\begin{tabular}{rll}
$m_R \;\; =$ & $(0.44624 \pm 0.00046 \pm 0.00007) \cdot a_t^{-1}$ & \multirow{2}{*}{ $\begin{bmatrix} 1 & 0.43 \\ & 1\end{bmatrix}$ } \\
$g_R \;\; =$ & $18.7 \pm 0.4 \pm 0.2$   & \\[1.3ex]
&\multicolumn{2}{l}{ $\chi^2/ N_\mathrm{dof} = \frac{13.4}{18 - 2} = 0.84\, . $}  \\[1.3ex]
\end{tabular}
\vspace{-0.75cm}
\end{center}
\begin{equation} \label{eq_D_bw_fit}\end{equation}

\subsection{Poles and Interpretation}\label{sec_poles}

Finite-volume energy levels determined from Euclidean two-point correlation functions are real and in the sections above we have used these energies to constrain the $t$-matrix at real values of energy. However, the $t$-matrix can also be considered a function of complex energies, where bound states and resonances can be associated with poles in the complex plane. In the proximity of a pole, $s_{\mathrm{pole}}$, the $t$-matrix is dominated by the term
\begin{equation}
t_{ij} \sim \frac{c_i c_j}{s_{\mathrm{pole}}-s}\, ,
\end{equation}
where the factorised residues, $c_{i}$, are complex numbers that give a measure of the ``coupling" of the pole to channel $i$.

In terms of complex energies, branch cuts appear in the $t$-matrix for each scattering threshold leading to $2^{N}$ Riemann sheets for $N$ coupled-channels. Sheets can be labelled by the sign of the imaginary part of the momenta, $k_{i}$, for each channel $i$. Poles that correspond to a resonance occur in complex conjugate pairs on ``unphysical sheets", where $\mathrm{Im}[k_{i}] < 0$ in at least one channel. The only poles permitted to occur on the ``physical sheet", where all $\mathrm{Im}[k_i] > 0$, are those corresponding to bound states. Bound states far below threshold are unlikely to influence physical scattering, but one sufficiently close to threshold can cause noticeable effects. We now proceed to interpret our results in terms of poles we find in our parametrised $t$-matrices.

\subsubsection{$S$-wave}

\begin{figure}[t!]
\begin{center}
\includegraphics[width=0.99\textwidth]{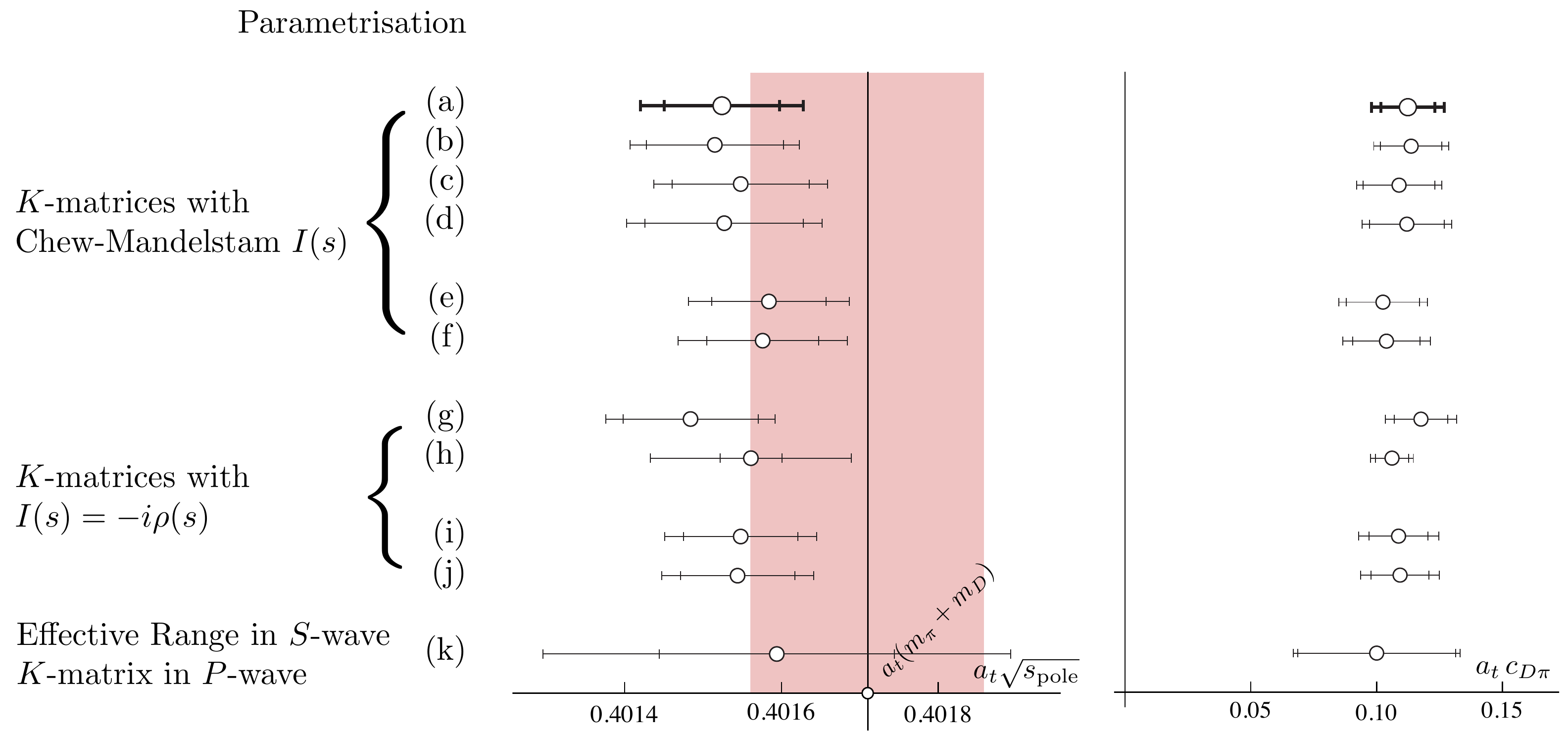}
\caption{The central (right) panel shows the positions of the poles (couplings) extracted from various parametrisations of elastic $D\pi$ scattering -- labels in the left panel refer to Table~\ref{tab_elastic_variations} which gives details of the parametrisations. The inner error bars show the one-sigma statistical uncertainty while the outer ones show the additional uncertainty coming from the scattered meson masses and the anisotropy. The red band represents the statistical uncertainty of the $D\pi$ threshold.}
\label{fig_elastic_S_pole}
\end{center}
\end{figure}

To begin, we investigate the pole structure of the $S$-wave parametrisations used to describe elastic $D\pi$ scattering in Section~\ref{sec:swave_elastic_param_vars}. In all of the parametrisations we considered, we consistently find a bound-state pole on the real axis of the physical sheet extremely close to the $D\pi$ threshold. In the central panel of Fig.~\ref{fig_elastic_S_pole}, we show the location of this pole for each parametrisation along with the $\chi^2/N_{\mathrm{dof}}$ in each case. By averaging over the pole positions from parametrisations with $\chi^2_\mathrm{dof} <1.9$, we find 
\begin{equation}
a_t \sqrt{s_{\mathrm{pole}}} = 0.40155 \pm 0.00015 ~,
\end{equation}
where the quoted uncertainty encompasses the uncertainties from the individual parametrisations.
Although the central value lies below the $D\pi$ threshold, which is located at $a_t E_\cm = 0.40171\pm 0.00015$, they overlap within uncertainties. The effect of the pole is seen in our $S$-wave amplitude as shown in Fig.~\ref{fig_elastic_S_kcot_amp_par_var}, where we observe a rapid variation coincident with the $D\pi$ threshold.
We also extract the residue of the pole, measuring the strength of its coupling to the $D\pi$ channel; this is shown in the right panel of Fig.~\ref{fig_elastic_S_pole}. Since the central value of the pole is below threshold, the residue has no imaginary part. We find that, like the pole, the residue is very stable across parametrisations, and averaging over all parametrisations shown we obtain 
\begin{equation}
a_t c_{D\pi} = 0.110 \pm 0.025\,.
\end{equation}
We do not find any further poles in the region where we have constrained the amplitudes. 

In all of the coupled-channel parametrisations used in Section~\ref{sec:swave_elastic_param_vars},  we find a pole consistent with that of the elastic case described above. Averaging over the coupled-channel parametrisations with a $\chi^2/N_{\mathrm{dof}} < 1.9$, we find our final location of the pole to be 
\begin{equation}
a_t \sqrt{s_{\mathrm{pole}}} = 0.40161 \pm 0.00015 \, .
\end{equation}
The pole couplings in the coupled-channel amplitudes are
\begin{equation}
a_t c_{\pD}  = 0.097 \pm 0.028 \, , \quad 
a_t c_{\eD}  = 0.077 \pm 0.023 \, , \quad 
a_t c_{\KDs} = 0.039 \pm 0.015 \,.
\end{equation}
The $\eD$ and $\KDs$ pole couplings involve a large analytic continuation from where they are kinematically open, and therefore constrained by the spectra, to the position of the pole. The successful parametrisations have similar properties and all used the Chew-Mandelstam phase space.

\subsubsection{$P$-wave}

In the $P$-wave part of the $t$-matrices determined in Section~\ref{sec:coupled_scat_swave}, we consistently find a bound-state pole on the real axis of the physical sheet well below threshold. Averaging over parametrisations with $\chi^2/N_{\mathrm{dof}} < 1.9$, we obtain our final position for the pole to be
\begin{equation}
a_t \sqrt{s_{\mathrm{pole}}} = 0.35440 \pm 0.00023 \, .
\end{equation}
As discussed in Section~\ref{sec:pwave_scat}, this pole can be associated with the stable $J = 1^{-}$ state found at almost exactly the same energy in Ref.~\cite{Moir:2013ub}. Because this state is far below the $D\pi$ threshold it is not expected to strongly influence $D\pi$ scattering; this is consistent with the small $P$-wave amplitudes we find.

\subsubsection{$D$-wave}

\begin{figure}[t!]
\begin{center}
\includegraphics[width=0.8\textwidth]{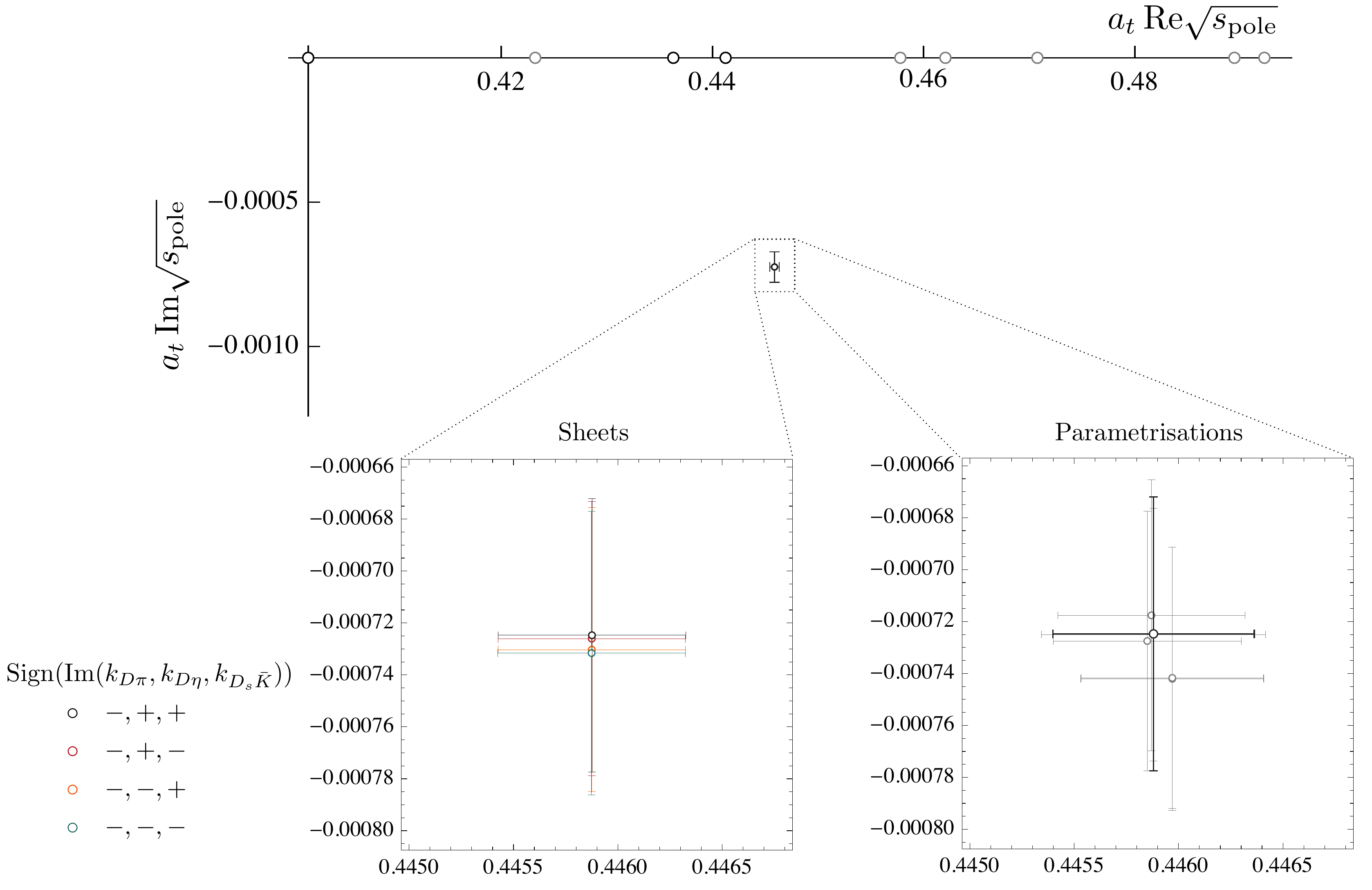}
\caption{The location of the pole found in our $D$-wave scattering amplitudes. The inner left panel shows its location on each of the Riemann sheets from the parametrisation in Eq.~\ref{fit_minimum_3x3_D}. The inner right panel shows its location on the $\mathrm{Sign}(\mathrm{Im}[k_{D\pi}, k_{D\eta}, k_{D_{s}\bar{K}}]) = (-,+,+)$ sheet for the parametrisations listed in Table~\ref{tab_D_wave_par_vars}.}
\label{fig_D_pole}
\end{center}
\end{figure}

In Section~\ref{sec:Dwave_param_vars}, we determined $D$-wave scattering amplitudes by considering the $D\pi$, $D\eta$ and $D_s\bar{K}$ channels\footnote{We reiterate that we did not consider the $D^\star \pi$ channel.}. In all of the parametrisations of the $t$-matrix we considered, we found what appeared to be an ``extra level" in close proximity to both the $D\eta$ and $D_{s}\bar{K}$ thresholds along with a rapid phase shift through $180^{\circ}$ in the $D\pi$ channel. We find an isolated resonance pole on each of the Riemann sheets with $\mathrm{Im}[k_{D\pi}] < 0$. As shown in Fig.~\ref{fig_D_pole}, very little variation in the pole position is observed across sheets or parametrisations. Averaging over all of our parametrisations, we find the final position of the pole to be
\begin{equation}
a_t \sqrt{s_{\mathrm{pole}}} = \left(0.44588\pm 0.00052\right) - \frac{i}{2}\left(0.00145\pm 0.00012\right) \, .
\end{equation} 

\begin{figure}[t!]
\begin{center}
\includegraphics[width=0.6\textwidth]{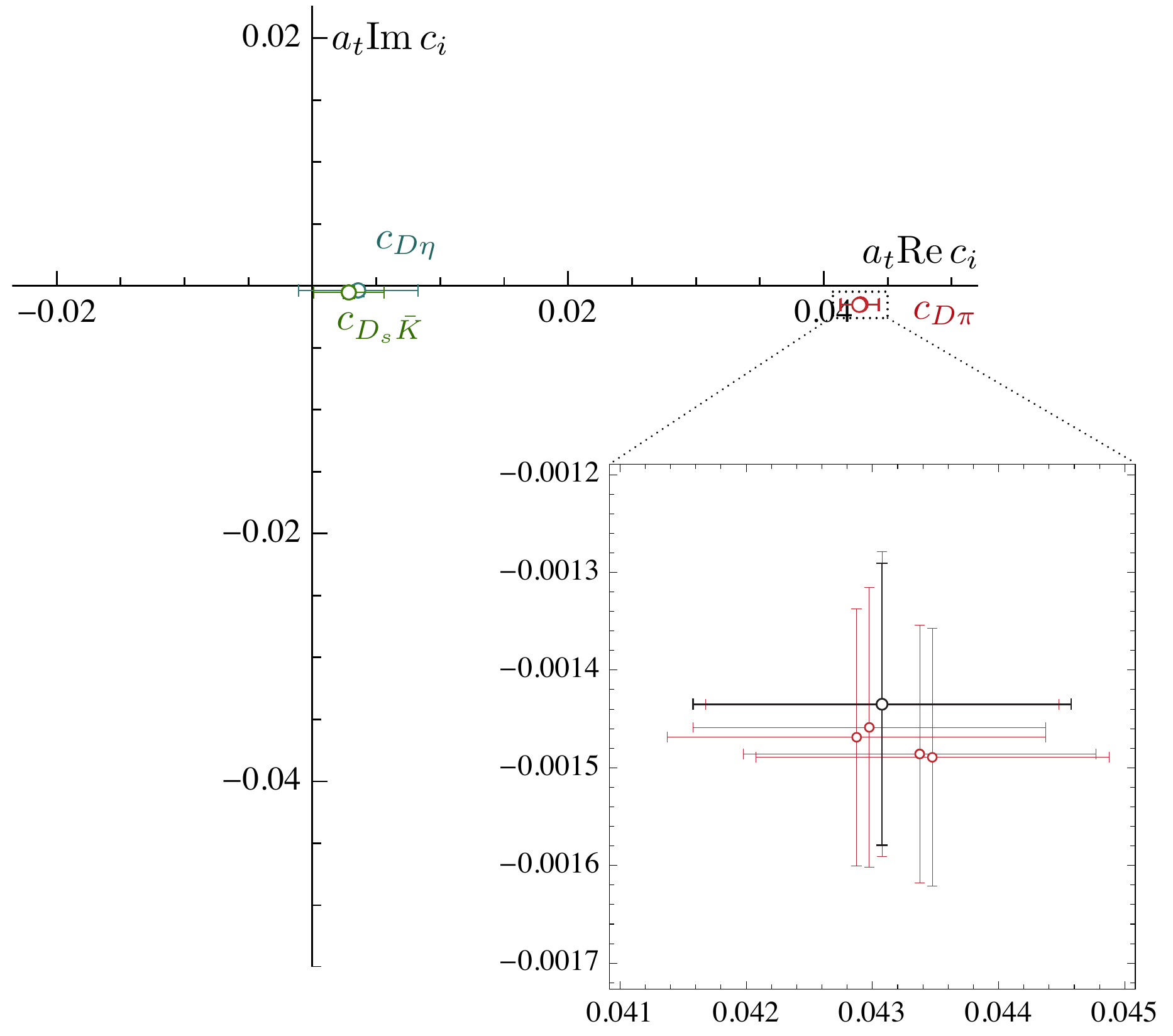}
\caption{The coupling, $c_i$, of the $D\pi$ (red), $D\eta$ (green) and $D_{s}\bar{K}$ (blue) channels to the $D$-wave resonance pole shown in Fig.~\ref{fig_D_pole}. The inset shows the coupling to the $D\pi$ channel for the various parametrisations listed in Table~\ref{tab_D_wave_par_vars} with the black point corresponding to the parametrisation in Eq.~(\ref{fit_minimum_3x3_D}).}
\label{fig_D_res}
\end{center}
\end{figure}

We also determine the couplings of each channel to the pole. As shown in Fig.~\ref{fig_D_res}, we find a non-zero coupling only in the $D\pi$ channel; as expected from Section~\ref{sec:Dwave_param_vars}, the $D\eta$ and $D_{s}\bar{K}$ channels are decoupled from the resonance. Averaging over all of our parametrisations, we obtain our final value for the coupling
\begin{equation}
a_t c_{D\pi} = (0.0431\pm 0.0015) \, \cdot \, \exp i\pi (-0.0106 \pm 0.0013) \, .
\end{equation}

\section{Results: Isospin-$3/2$}\label{sec:iso_three_half}

We now change focus and present the results of elastic $D\pi$ scattering in the isospin-$3/2$ channel. In our calculation we include only $D\pi$ interpolating operators: there are no $\bar{q}q$ operators with this isospin.

\subsection{Finite-Volume Spectra}\label{subsec:iso_three_half_spectra}

\begin{figure}[t!]
\begin{center}
\includegraphics[width=0.95\textwidth]{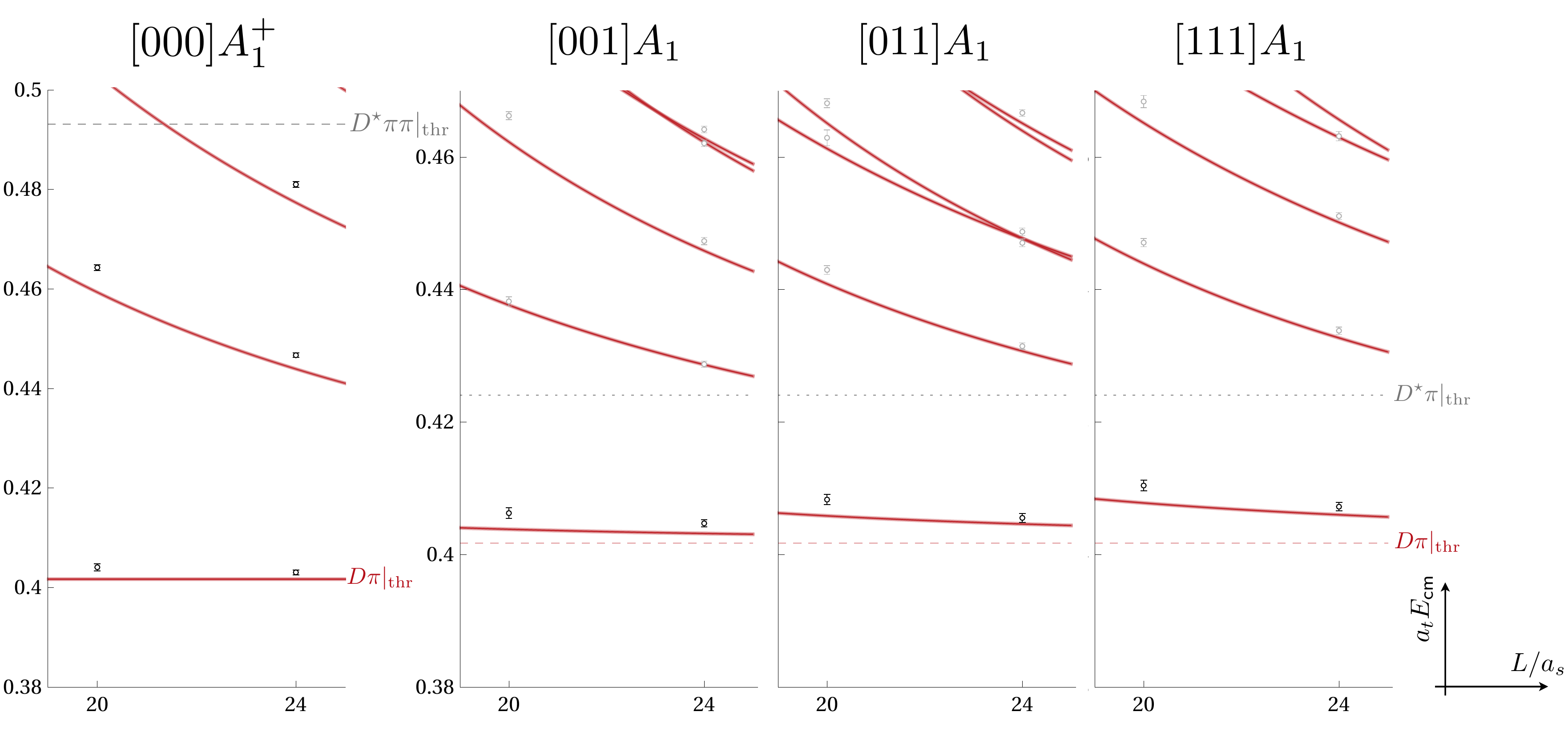}
\end{center}
\caption{Finite-volume isospin-$3/2$ spectra obtained from the $[000]A^{+}_{1}, [001]A_{1}, [011]A_{1}$ and $[111]A_{1}$ irreps. The black (grey) points correspond to energy levels that we use (do not use) in our subsequent scattering analysis. The solid curves represent non-interacting $D\pi$ energies while the red (grey) dashed lines correspond to a channel threshold for which we have (have not) included operators in our variational procedure.}
\label{fig:I32_A1_spectra}
\end{figure}

As before, we determine energy levels from a variational procedure applied to a matrix of two-point correlation functions. We construct these correlation functions using the $D\pi$ operators listed in Tables~\ref{tab_I32_ops_A1} and~\ref{tab_I32_ops_P} in Appendix \ref{app:op_tables}. In Fig.~\ref{fig:I32_A1_spectra} we show our finite-volume spectra in irreps that have $\ell = 0$ as the lowest contributing partial wave, namely $[000]A^{+}_{1}, [001]A_{1}, [011]A_{1}$ and $[111]A_{1}$. In all four spectra, we observe small positive shifts of our energy levels from the non-interacting energies, which is usually indicative of a weakly repulsive interaction. 

\begin{figure}[t!]
\begin{center}
\includegraphics[width=0.95\textwidth]{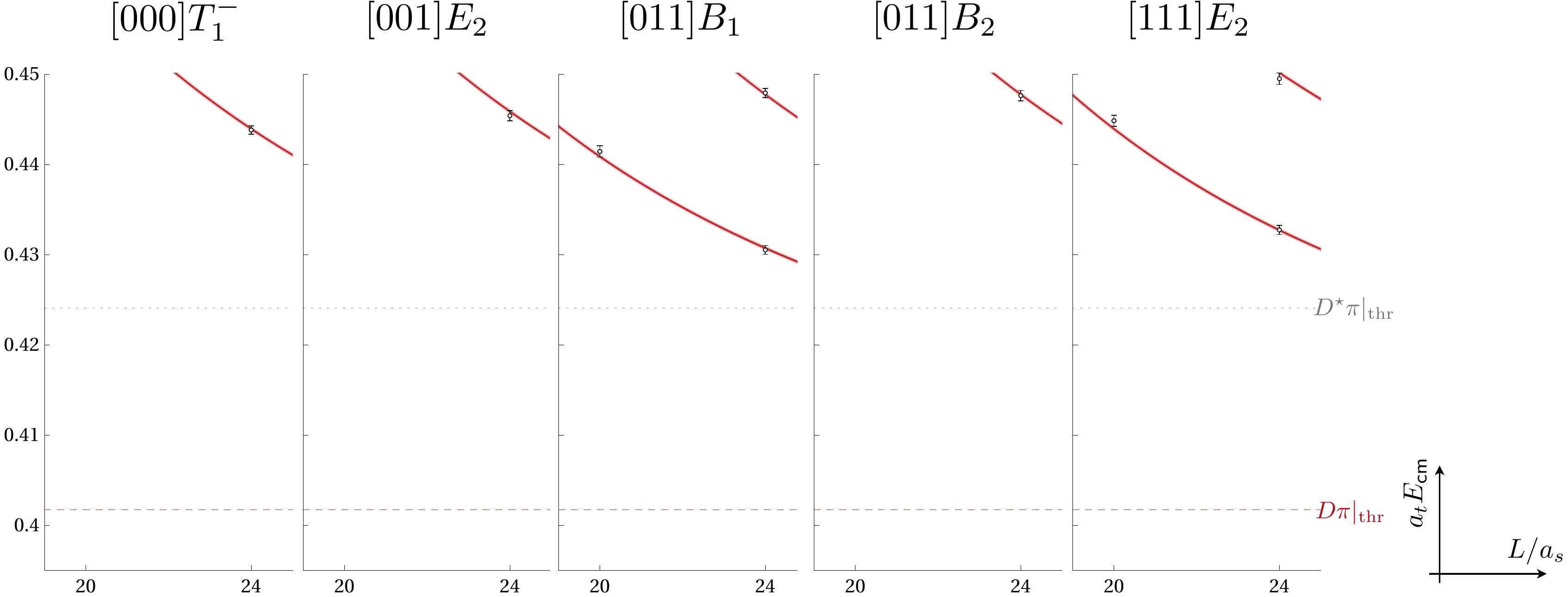}
\end{center}
\caption{As Fig. \ref{fig:I32_A1_spectra} but for the $[000]T_{1}^{-}, [001]E_{2}, [011]B_{1}, [011]B_{2}$ and $[111]E_{2}$ irreps.}
\label{fig:I32_nonA1_spectra}
\end{figure}

In Fig.~\ref{fig:I32_nonA1_spectra}, we show our determined energy levels for irreps that have $\ell = 1$ as the lowest contributing partial wave, namely $[000]T_{1}^{-}, [001]E_{2}, [011]B_{1}, [011]B_{2}$ and $[111]E_{2}$.

\subsection{Elastic $D\pi$ Scattering}\label{subsec:iso_three_half_s_wave}

\begin{figure}[t!]
\begin{center}
\includegraphics[width=0.95\textwidth]{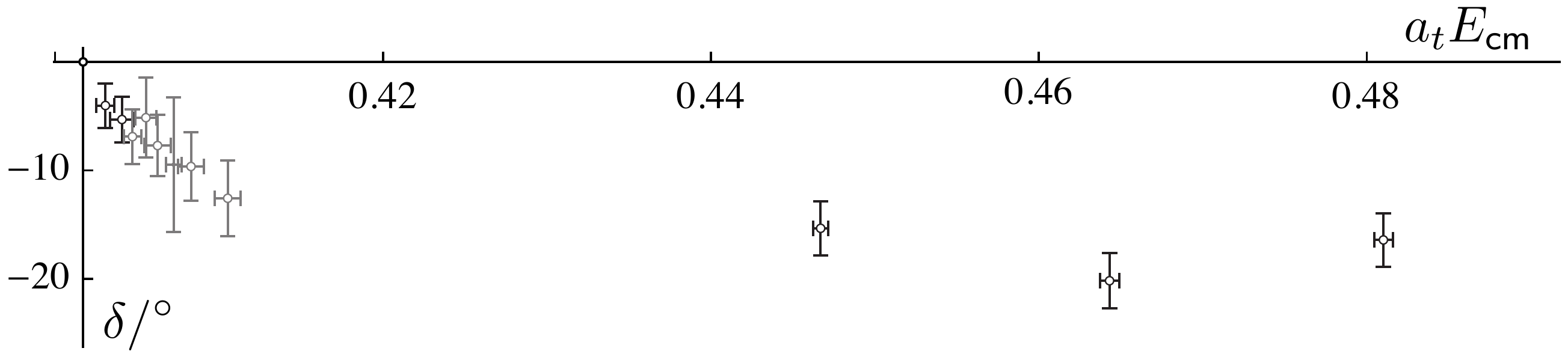}
\end{center}
\caption{The $S$-wave phase shift $\delta_{0}$ for $D\pi$ scattering in the isospin-$3/2$ channel. The black points are determined using the $[000]A^{+}_{1}$ spectrum of Fig.~\ref{fig:I32_A1_spectra}, while the grey points are determined using the $[001]A_{1}, [011]A_{1}$ and $[111]A_{1}$ spectra of Fig.~\ref{fig:I32_A1_spectra}.}
\label{fig:I32_discrete_phase_shift}
\end{figure}

We begin by using the energy levels coloured black in Fig.~\ref{fig:I32_A1_spectra} to map out the $S$-wave phase shift $\delta_{0}$ as a function of energy. To do so, we ignore higher partial waves and solve Eq.~(\ref{eqn:luscher}) for each considered energy level, resulting in a corresponding value for $\delta_{0}$ at that energy. We show the outcome of this procedure in Fig.~\ref{fig:I32_discrete_phase_shift}, where the black points represent the values of $\delta_{0}$ that have come from the $[000]A^{+}_{1}$ irrep and the grey points represent values that come from the $[001]A_{1}, [011]A_{1}$ and $[111]A_{1}$ irreps. The form of the phase shift is consistent with that of a weakly repulsive interaction, which may be expected in the isospin-$3/2$ channel. As mentioned above, we have neglected contributions of higher partial waves; in determining the black points we have neglected contributions coming from $\ell \geq 4$, which is justified due to the large angular momentum suppression, and in determining the grey points we have neglected contributions coming from $\ell \geq 1$. To justify the use of the grey points, we assume negligible inelasticity into $D^{\star}\pi$ and compute the magnitude of the $P$-wave phase shift, $\delta_{1}$, within the energy range $0.43 \leq a_{t}E_{\cm} \leq 0.45$ using the energy levels in Fig.~\ref{fig:I32_nonA1_spectra}. We find that $|\delta_{1}| \leq 5^{\circ}$ at $a_{t}E_{\cm} \approx 0.45$ and $|\delta_{1}| \leq 3^{\circ}$ at $a_{t}E_{\cm} \approx 0.43$. We expect this trend to continue down to $a_{t}E_{\cm} \approx 0.41$, justifying our earlier assumption. \\

\begin{figure}[t!]
\begin{center}
\includegraphics[width=0.6\textwidth]{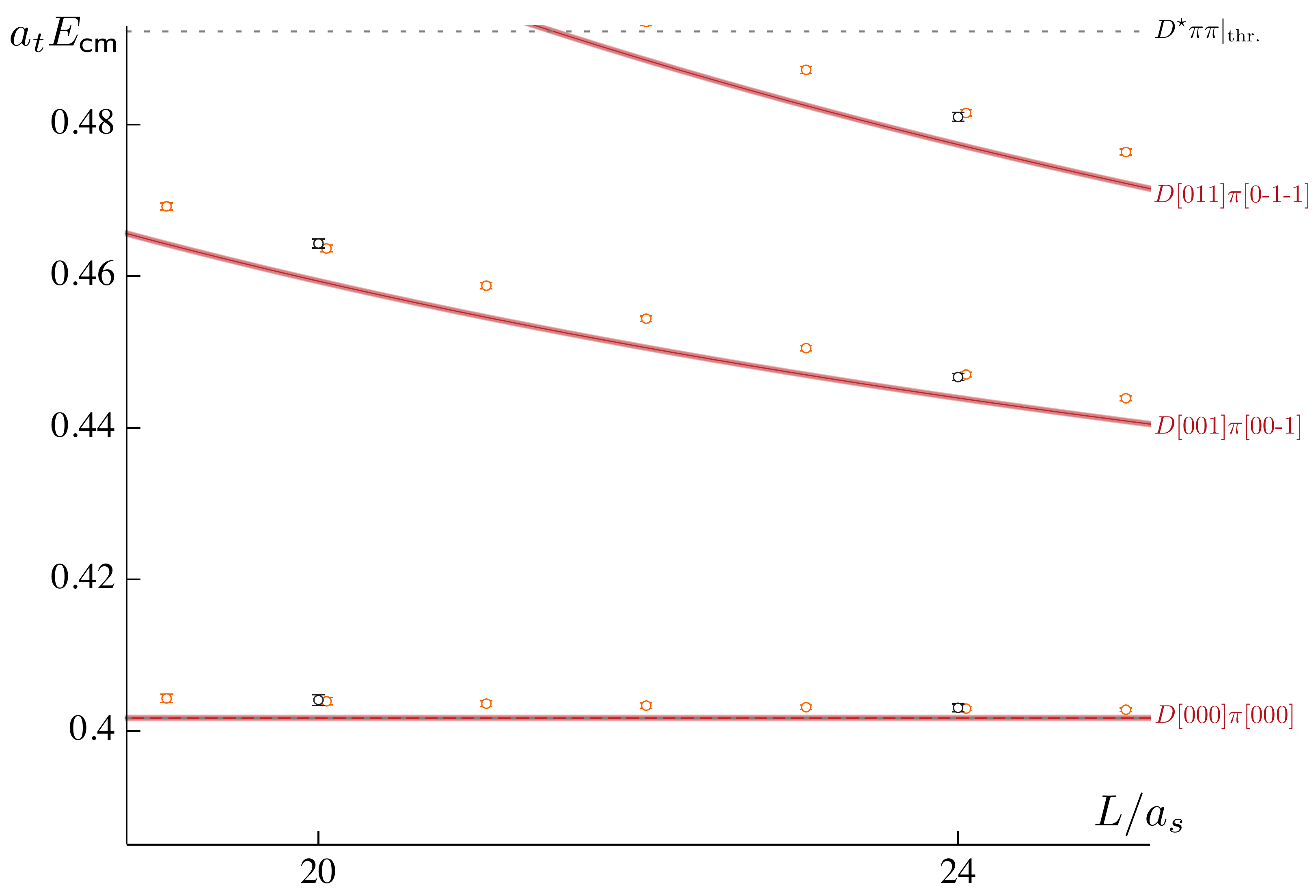}
\end{center}
\caption{As the left panel of Fig.~\ref{fig:I32_A1_spectra} but with the addition of orange points corresponding to the spectrum resulting from the parametrisation in Eq.~(\ref{eqn:I32_A1p_fit}).}
\label{fig:A1p_with_fit}
\end{figure}

We can also follow the approach taken in the isospin-$1/2$ section and use the determined spectra to constrain the scattering amplitude as a function of energy. By again ignoring the negligible contribution
coming from $\ell \geq 4$, we use the $[000]A_{1}^{+}$ energy levels in Fig.~\ref{fig:I32_A1_spectra} and parametrise the $t$-matrix using a scattering length and effective range. We find that the parameters
\begin{center}
\begin{tabular}{rll}
$a_{0} \;\; =$ & $(-5.4 \pm 1.0 \pm 0.3) \cdot a_{t}$ & \multirow{2}{*}{ $\begin{bmatrix} 1 & 0.52 \\ & 1\end{bmatrix}$ } \\
$r_{0} \;\; =$ & $(-25 \pm 6 \pm 6) \cdot a_{t}$   & \\[1.3ex]
&\multicolumn{2}{l}{ $\chi^2/ N_\mathrm{dof} = \frac{3.86}{5 - 2} = 1.29\, $}  \\
\end{tabular}
\end{center}
\vspace{-0.75cm}
\begin{equation} \label{eqn:I32_A1p_fit}\end{equation}
are sufficient to describe our determined spectrum. In Fig.~\ref{fig:A1p_with_fit} we show the comparison between our determined spectrum and the spectrum resulting from the parameters in Eq.~(\ref{eqn:I32_A1p_fit}). 

\begin{figure}[t!]
\begin{center}
\includegraphics[width=0.95\textwidth]{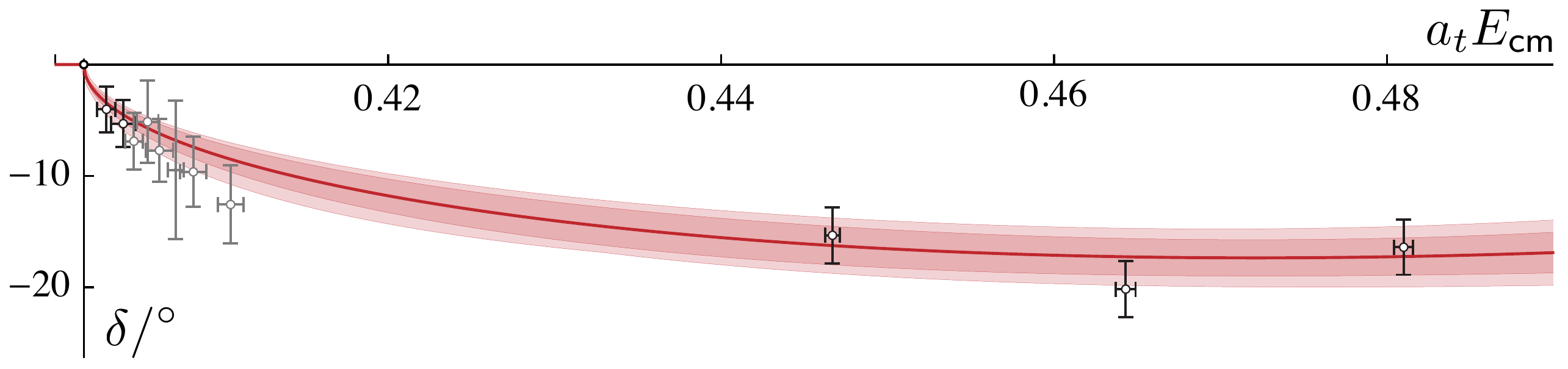}
\end{center}
\caption{The $S$-wave phase shift, $\delta_{0}$. The extent of the band encompasses the phase shifts resulting from the parametrisations with $\chi^2/N_\mathrm{dof} < 1.9$ in Table~\ref{tab:I32_vars} as well as the
systematic uncertainty coming from the anisotropy and the scattered meson masses. The points from Fig.~\ref{fig:I32_discrete_phase_shift} are superimposed for comparison.}
\label{fig:I32_phase_vars}
\end{figure}

We repeat this process for various parametrisations, and in Table \ref{tab:I32_vars} show a selection that were found to obtain a reasonable $\chi^2/N_\mathrm{dof}$. Our results suggest that the best description of our finite-volume spectrum is achieved using two free parameters. Fig.~\ref{fig:I32_phase_vars} shows the $S$-wave phase shift, $\delta_{0}$, where the size of the band includes parametrisations which were found to result in a $\chi^2/N_\mathrm{dof} < 1.9$. The points in Fig.~\ref{fig:I32_phase_vars} are taken from Fig.~\ref{fig:I32_discrete_phase_shift} and superimposed to demonstrate the consistency between this approach and the approach discussed above. 

\begin{table}[t!]
\begin{center}
\begin{tabular}{lcc}

Parametrisation                                                   & $N_\mathrm{pars}$ & $\chi^2/N_\mathrm{dof}$\\
\hline
K-matrix with Chew-Mandelstam $I(s)$ & &\\
$K=\gamma^{(0)}$                                                  & 1 & \textit{3.06} \\
$K=\gamma^{(0)} + \gamma^{(1)} s$                                 & 2 & 1.30 \\[0.4ex]
\hline
Effective range expansion & & \\
$k\cot\delta = \frac{1}{a}$                                       & 1 & \textit{3.66}\\
$k\cot\delta = \frac{1}{a}+\frac{1}{2}r^2 k^2$                    & 2 & 1.29 \\[0.4ex]
\hline
\end{tabular}
\end{center}
\caption{A selection of $S$-wave parametrisations for elastic isospin-$3/2$ $D\pi$ scattering. Only those with $\chi^2/N_\mathrm{dof} < 1.9$ are used in Fig.~\ref{fig:I32_phase_vars}.}
\label{tab:I32_vars}
\end{table}

To get a handle on the $P$-wave amplitudes we can include the energy levels coloured black from the $[\vec{P} \neq 0]A_{1}$ irreps in Fig.~\ref{fig:I32_A1_spectra}. However, we find that for all our forms of the $t$-matrix, the $P$-wave parameters are always consistent with zero. 
Furthermore, by setting the $P$-wave parameters to zero, we we see no significant variation in any of our $S$-wave parameters when we include energy levels coming from the $[\vec{P} \neq 0]A_{1}$ irreps. This agrees with what we found using the approach discussed above; the contribution coming from $\ell \geq 1$ below the $D^{\star}\pi$ threshold is negligible.

In physical units our result for the scattering length is $a_{0} = -0.19 \pm 0.05$ fm.  This value is in agreement with the results of Ref.~\cite{Liu:2012zya} where a chiral unitary approach is used to interpolate between different values of the pion mass.  

\section{Summary and Outlook}
\label{sec:conclusions}

\begin{figure}[t!]
\begin{center}
\includegraphics[width=0.99\textwidth]{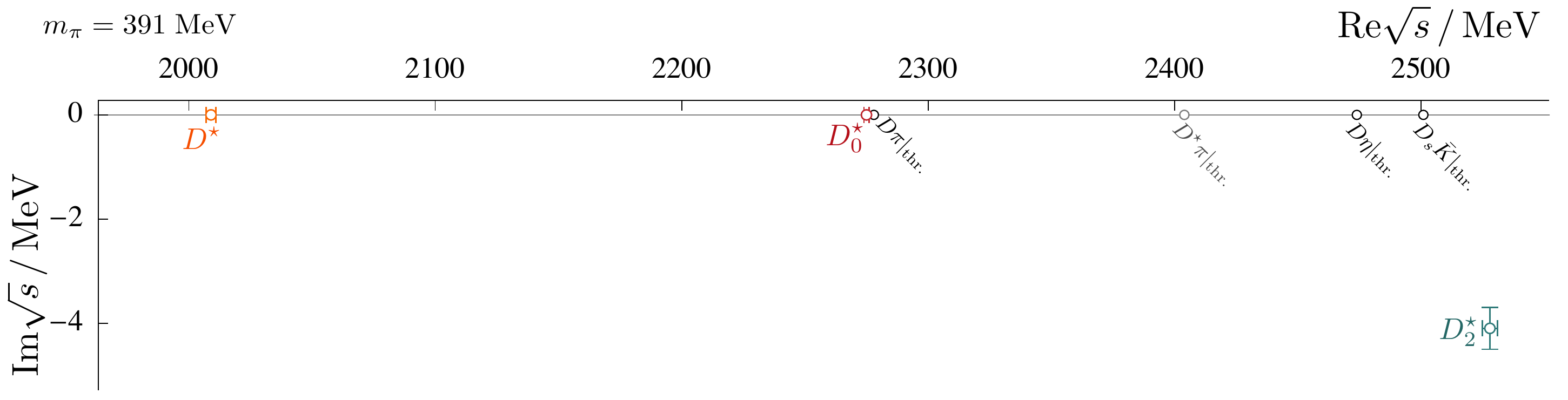}
\caption{The locations of the poles in the complex energy plane determined from coupled-channel isospin-$1/2$ $S$ ($D^{\star}_{0}$), $P$ ($D^{\star}$) and $D$-wave ($D^{\star}_{2}$) amplitudes. The black circles on the real axis correspond to relevant thresholds.}
\label{phys_pole}
\end{center}
\end{figure}

We have presented the first lattice QCD study of coupled-channel $D\pi$, $D\eta$ and $D_{s}\bar{K}$ scattering. Utilising the distillation framework and large bases of $\bar{q}q$, $D\pi$, $D\eta$ and $D_{s}\bar{K}$-like interpolating operators, we have determined finite-volume energy levels in many lattice symmetry channels.  These were used to constrain scattering amplitudes as a function of energy which we analytically continue to complex energies.  We determined the pole structure of these amplitudes and, from their residues, the coupling of poles to channels.  Figure~\ref{phys_pole} shows the resulting $S$, $P$ and $D$-wave isospin-$1/2$ poles.

\begin{figure}[t!]
\begin{center}
\includegraphics[width=0.8\textwidth]{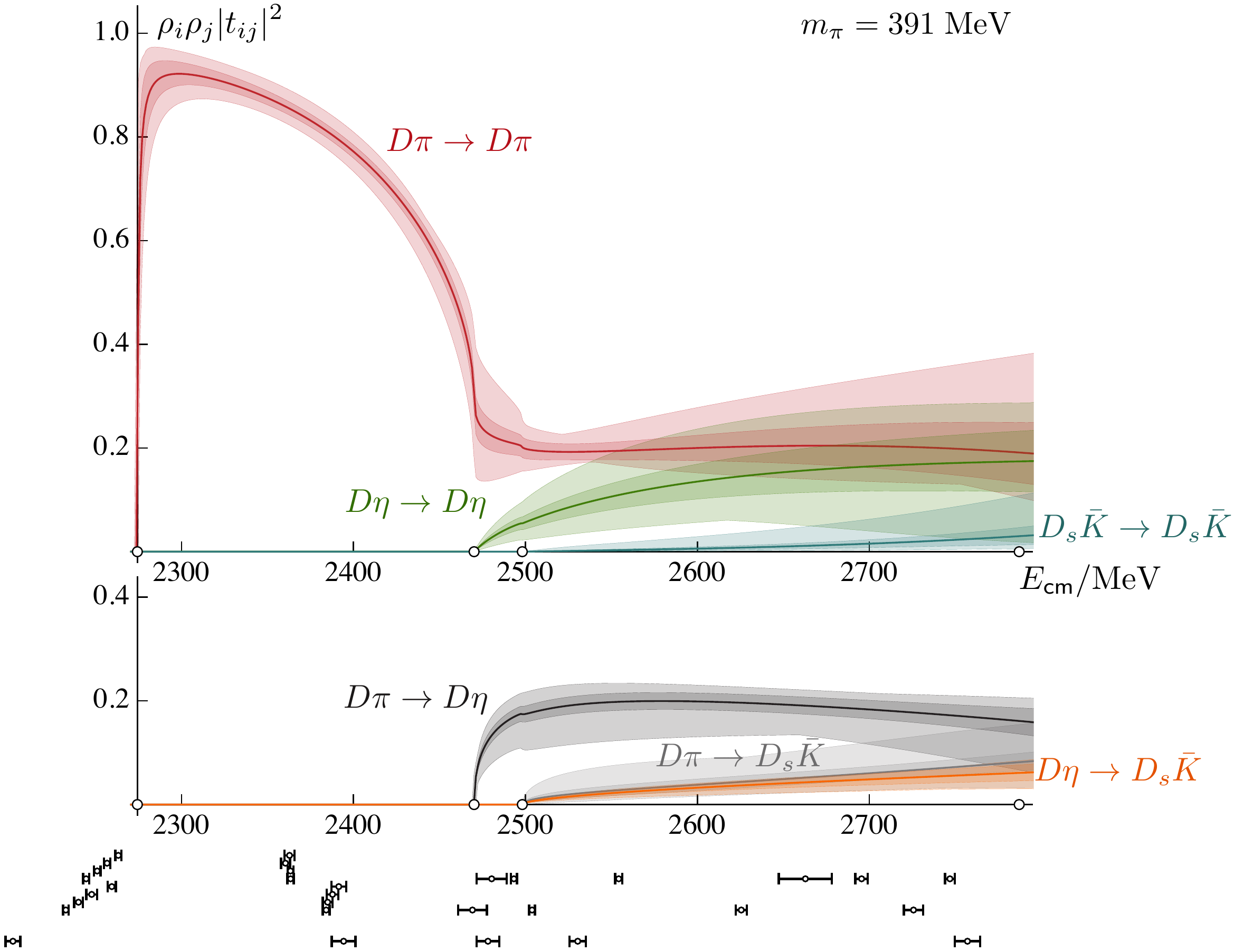}
\caption{$\rho_{i}\rho_{j}|t_{ij}|^{2}$ for $S$-wave scattering in the isospin-$1/2$ channel.  The bands encompass all the parametrisations with $\chi^2/N_\mathrm{dof} < 1.9$ in Table~\ref{tab_coupled_variations} along with the uncertainties coming from the scattered meson masses and the anisotropy.  Black points show the location of the finite-volume energy levels used to constrain the scattering amplitudes.}
\label{phys_amps}
\end{center}
\end{figure}

In the $J^{P} = 0^{+}$ channel, we find a pole just below the $D\pi$ threshold signalling the presence of a bound state. Although this pole appears to couple predominantly to $D\pi$, we obtain significant couplings to the $D\eta$ and $D_{s}\bar{K}$ channels. As shown in Fig.~\ref{phys_pole}, we find the pole to be at $(2275.9 \pm 0.9)$ MeV, which is statistically indistinguishable from the $D\pi$ threshold which is at $(2276.4 \pm 0.9)$ MeV on our ensembles. As a consequence, the effect of the pole can be seen above threshold; as shown in Fig.~\ref{phys_amps}, we observe a large ``peak" in $\rho_{D\pi}^{2}|t_{D\pi,D\pi}|^{2}$ almost saturating the unitarity bound. Since $m_{\pi} = 391$ MeV in this calculation, we only make a qualitative comparison with experiment. Although we find a near-threshold bound state, it shares similarities with the experimental $D^{\star}_{0}(2400)$ resonance; both states couple dominantly to $D\pi$ and influence a similar broad energy range \cite{Agashe:2014kda}.  Noting that the light quark mass in this calculation lies between the physical light and strange quark masses, and ignoring the differences due to flavour, the relative position of the pole to the threshold lies between what is observed experimentally for the resonant $D^{\star}_{0}(2400)$ and the bound $D_{s0}^{\star}(2317)$.  Qualitatively, such behaviour is anticipated from unitarised chiral perturbation theory amplitudes~\cite{Guo:2009ct, Liu:2012zya, Guo:2015dha}.

As shown in Fig.~\ref{phys_pole}, we find a pole at $(2009 \pm 2)$ MeV in the $1^{-}$ channel corresponding to a deeply-bound state, consistent with what was found in Ref.~\cite{Moir:2013ub}. Experimentally, the near-threshold $D^{\star}(2007)$ resonance, which is narrow and decays predominantly to $D\pi$, has a mass of $(2006.97 \pm 0.08)$ MeV \cite{Agashe:2014kda}.

In the $2^{+}$ channel we find a narrow resonance coupled to $D\pi$. As shown in Fig.~\ref{phys_pole}, we determine its pole mass and width to be $(2527 \pm 3)$ MeV and $(8.2 \pm 0.7)$ MeV respectively. Experiment finds a relatively narrow tensor resonance, the $D^{\star}_{2}(2460)$, coupled to $D\pi$. However, this also couples to $D^{\star}\pi$, a kinematically open channel which we have neglected in the determination of our $2^{+}$ state.

We have also performed a study of elastic $D\pi$ scattering in the isospin-$3/2$ channel. We find that the weakly-repulsive $S$-wave interaction can be successfully described using a scattering length and effective range parametrisation with $a_{0} = -0.19 \pm 0.05$ fm and $r_{0} = -0.9 \pm 0.4$ fm.

This work, which is the first \textit{ab initio} coupled-channel study including charm quarks, has taken a significant step towards understanding the striking differences between the $D^{\star}_{0}(2400)$ and $D^{\star}_{s0}(2317)$.  A complementary study of $DK$ scattering is already underway and in the near future we will perform calculations with lighter pion masses -- these will enable a more direct comparison with experiment and allow us to study the dependence of the pole position on the light-quark mass. These calculations will also include additional channels, such as $D^{\star}\pi$, whose role may become increasingly important as the light quark mass is reduced.

\bigskip

\begin{acknowledgments}
We thank our colleagues within the Hadron Spectrum Collaboration. GM acknowledges support from the Herchel Smith
Fund at the University of Cambridge and the Deutsche Forschungsgemeinschaft (DFG) under contract KN 947/1-2.
SMR acknowledges support from Science Foundation Ireland [RFP-PHY-3201].
CET acknowledges support from the U.K. Science and Technology Facilities Council (STFC) [grant ST/L000385/1]
and the Isaac Newton Trust/University of Cambridge Early Career Support Scheme [RG74916].

The software codes {\tt Chroma}~\cite{Edwards:2004sx} and {\tt QUDA}~\cite{Clark:2009wm,Babich:2010mu} were used to perform this work at Jefferson Laboratory under the USQCD Initiative and the LQCD ARRA project, and on the Lonsdale cluster maintained by the Trinity Centre for High Performance Computing funded through grants from Science Foundation Ireland (SFI). This work also used the DiRAC Data Analytic system at the University of Cambridge, operated by the University of Cambridge High Performance Computing Service on behalf of the STFC DiRAC HPC Facility (www.dirac.ac.uk). This equipment was funded by BIS National E-infrastructure capital grant ST/K001590/1, STFC capital grants ST/H008861/1 and ST/H00887X/1, and STFC DiRAC Operations grant ST/K00333X/1. DiRAC is part of the National E-Infrastructure. This research also used the Wilkes GPU cluster at the University of Cambridge High Performance Computing Service (http://www.hpc.cam.ac.uk/), provided by Dell Inc., NVIDIA and Mellanox, and part funded by STFC with industrial sponsorship from Rolls Royce and Mitsubishi Heavy Industries. Gauge configurations were generated using resources awarded from the U.S. Department of Energy INCITE program at Oak Ridge National Laboratory, the NSF Teragrid at the Texas Advanced Computer Center and the Pittsburgh Supercomputer Center, as well as at Jefferson Lab.
\end{acknowledgments}

%\newpage

\appendix
\section*{Appendices}
\section{Lattice Irreps and Partial Waves}\label{app:tables}

In Table~\ref{tab:subductions} we show the pattern of contributing partial waves, $\ell$, for various overall momentum types, $\vec{P}$, and lattice irreps, $\Lambda$, relevant for the scattering of two unequal-mass pseudoscalars.  Note that, because the mesons have different masses, even and odd partial waves mix for non-zero overall momentum.

\begin{table}[t!]
  \begin{center}
    \begin{tabular}{cc|c|l|l}
           &&&&\\[-1.7ex]
      \multirow{2}{*}{$\vec{P}$} & \multirow{2}{*}{LG$(\vec{P})  $ }  & \;  \multirow{2}{*}{$\Lambda$} \;& $\,\,J^P (\vec{P}=\vec{0})$    & \; \multirow{2}{*}{$\ell^N$} \\  
      &               &       & $\left|\lambda\right|^{({\tilde{\eta}})}(\vec{P}\neq\vec{0})$ & \\[0.5ex]
      \hline \hline
           &&&&\\[-1.5ex]
      \multirow{7}{*}{$\left[0,0,0\right]$}&     \multirow{7}{*}{$\textrm{O}_h^\textrm{D}$ ($\textrm{O}_h$)} & $A_1^+$            
      & $0^+,\, 4^+$          &\; $0^1,\, 4^1$\\
      && $T_1^-$    & $1^-,\, 3^-,\, \mathit{(4^-)}$ &\; $1^1,\, 3^1$\\
      && $E^+$      & $2^+,\, 4^+$          &\; $2^1,\, 4^1$\\
      && $T_2^+$    & $2^+,\, 4^+,\,  \mathit{(3^+)}\,$  &\; $2^1,\, 4^1$\\
      && $T_1^+$    & $4^+, \, \mathit{(1^+,3^+)}$ &\; $4^1$ \\
      && $T_2^-$    & $3^-,\, \mathit{(2^-,4^-)}$ &\; $3^1$ \\
      && $A_2^-$    & $3^-$               &\; $3^1$\\[0.5ex]
      \hline
      \hline 
      &&&&\\[-1.5ex]
      
      \multirow{5}{*}{$\left[0,0,n\right]$} & \multirow{5}{*}{Dic$_4$ ($\textrm{C}_{4\textrm{v}}$)}
      & $A_1$      & $0^+,\, 4$        &\; $0^1,\, 1^1,\, 2^1,\, 3^1,\, 4^2$ \\
      && $E_2$      & $1,\, 3$          &\; $1^1,\, 2^1,\, 3^2,\, 4^2$\\
      && $B_1$      & $2$               &\; $2^1,\, 3^1,\, 4^1$ \\
      && $B_2$      & $2$               &\; $2^1,\, 3^1,\, 4^1$ \\ 
      && $A_2$      & $4,\, \mathit{(0^-)}$      &\; $4^1$ \\[0.5ex]    
      \hline
      &&&&\\[-1.5ex]
      
      \multirow{4}{*}{$\left[0,n,n\right]$} & \multirow{4}{*}{Dic$_2$ ($\textrm{C}_{2\textrm{v}}$)} 
      & $A_1$     & $0^+,\, 2,\, 4$   &\; $0^1,\, 1^1,\, 2^2,\, 3^2,\, 4^3$ \\
      && $B_1$     & $1,\, 3$          &\; $1^1,\, 2^1,\, 3^2,\, 4^2$ \\
      && $B_2$     & $1,\, 3$          &\; $1^1,\, 2^1,\, 3^2,\, 4^2$ \\
      && $A_2$     & $2,\, 4, \, \mathit{(0^-)}$ &\; $2^1,\, 3^1,\, 4^2$ \\[0.5ex]      
      \hline
      &&&&\\[-1.5ex]
   
      \multirow{3}{*}{$\left[n,n,n\right]$} &       \multirow{3}{*}{Dic$_3$ ($\textrm{C}_{3\textrm{v}}$)}
      & $A_1$     & $0^+,\, 3$        &\;  $0^1,\, 1^1,\, 2^1,\, 3^2,\,  4^2$\\
      && $E_2$     & $1,\, 2,\, 4$     &\;  $1^1,\, 2^2,\, 3^2,\,  4^3$ \\
      && $A_2$     & $3, \, \mathit{(0^-)}$      &\;  $3^1,\, 4^1$\\[0.5ex]
      \hline
    \end{tabular}
  \end{center}
\caption{The pattern of subductions of pseudoscalar-pseudoscalar partial-waves, $\ell \leq 4$, into lattice irreps, $\Lambda$, when the pseudoscalars have unequal mass, e.g.\ $D\pi$, $D\eta$ or $D_{s}\bar{K}$ (from Table III of Ref.~\cite{Wilson:2014cna}).
Here $N$ is the number of embeddings of this $\ell$ in the irrep and $n$ is a non-zero integer. LG$(\vec{P})$ is the double-cover little group and the corresponding single-cover little group relevant for only integer spin is given in parentheses.  Also shown are the various $J \leq 4$ or $|\lambda| \leq 4$ that appear in each of the relevant irreps. The $J^P$ values and $|\lambda|^{\tilde{\eta}}=0^-$ in italics are in the ``unnatural parity'' [$P = (-1)^{J+1}$] series and do not contribute to pseudoscalar-pseudoscalar scattering.}
\label{tab:subductions}
\end{table}

\section{Operator Lists}\label{app:op_tables}

In Tables~\ref{tab_ops_A1},~\ref{tab_ops_P} and~\ref{tab_ops_D} we list the interpolating operators used to determine the finite-volume energy levels in the isospin-$1/2$ channel shown in Figs.~\ref{spec_A1},~\ref{spec_P} and~\ref{spec_D} respectively.
Tables~\ref{tab_I32_ops_A1} and~\ref{tab_I32_ops_P} show the operators used in the isospin-$3/2$ channel to determine the finite-volume spectra shown in Figs.~\ref{fig:I32_A1_spectra} and~\ref{fig:I32_nonA1_spectra} respectively; note that there are no $\bar{q}q$ operators with this isospin.

\begin{table}
\begin{center}
\begin{tabular}{c|c|c|c}
\multicolumn{1}{c|}{$[000]A_1^+$}      &\multicolumn{1}{c|}{$[001]A_1$} & \multicolumn{1}{c|}{$[011]A_1$} & \multicolumn{1}{c}{$[111]A_1$} \\
\hline \hline
$D_{000}\pi_{000}$                     & $D_{001}\pi_{000}$             & $D_{011}\pi_{000}$              & $D_{111}\pi_{000}$ \\  
$D_{001}\pi_{00\sm1}$                  & $D_{000}\pi_{00\sm1}$          & $D_{000}\pi_{0\sm1\sm1}$        & $D_{000}\pi_{\sm1\sm1\sm1}$ \\
$D_{011}\pi_{0\sm1\sm1}$               & $D_{011}\pi_{00\sm1}$          & $D_{001}\pi_{0\sm10}$           & $D_{011}\pi_{\sm100}$  \\
$^*D_{111}\pi_{\sm1\sm1\sm1}$          & $D_{001}\pi_{0\sm1\sm1}$       & $D_{011}\pi_{\sm1\sm10}$        & $D_{001}\pi_{\sm1\sm10}$  \\
$D_{000}\eta_{000}$                    & $D_{111}\pi_{0\sm1\sm1}$       & $D_{111}\pi_{00\sm1}$           & $D_{002}\pi_{\sm1\sm1\sm1}$ \\
$D_{001}\eta_{00\sm1}$                 & $D_{011}\pi_{\sm1\sm1\sm1}$    & $D_{001}\pi_{\sm1\sm1\sm1}$     & $D_{111}\pi_{00\sm2}$ \\
$^*D_{011}\eta_{0\sm1\sm1}$            & $D_{002}\pi_{00\sm1}$          & $D_{002}\pi_{0\sm1\sm1}$        & $D_{111}\eta_{000}$ \\
$D_{s\,000}\bar{K}_{000}$              & $D_{001}\eta_{000}$            & $D_{011}\eta_{000}$             & $D_{000}\eta_{\sm1\sm1\sm1}$\\
$D_{s\,001}\bar{K}_{00\sm1}$           & $D_{000}\eta_{00\sm1}$         & $D_{000}\eta_{0\sm1\sm1}$       & $D_{011}\eta_{\sm100}$ \\
$^*D_{s\,011}\bar{K}_{0\sm1\sm1}$      & $D_{011}\eta_{00\sm1}$         & $D_{001}\eta_{0\sm10}$          & $D_{001}\eta_{\sm1\sm10}$ \\
                                       & $D_{001}\eta_{0\sm1\sm1}$      & $D_{111}\eta_{00\sm1}$          & $D_{s\,111}\bar{K}_{000}$ \\
                                       & $D_{002}\eta_{00\sm1}$         & $D_{s\,011}\bar{K}_{000}$       & $D_{s\,000}\bar{K}_{\sm1\sm1\sm1}$\\
                                       & $D_{s\,001}\bar{K}_{000}$      & $D_{s\,000}\bar{K}_{0\sm1\sm1}$ & $D_{s\,011}\bar{K}_{\sm100}$ \\
                                       & $D_{s\,000}\bar{K}_{00\sm1}$   & $D_{s\,001}\bar{K}_{0\sm10}$    & $D_{s\,001}\bar{K}_{\sm1\sm10}$\\
                                       & $D_{s\,011}\bar{K}_{00\sm1}$   & $D_{s\,111}\bar{K}_{00\sm1}$    & \\
                                       &$D_{s\,001}\bar{K}_{0\sm1\sm1}$ &                                 & \\[1ex]

$(\singleop) \times 11$                & $(\singleop) \times 32$        & $(\singleop) \times 52$      & $(\singleop) \times 37$ \\
\hline
\end{tabular}
\end{center}
\caption{The interpolating operators used in each irrep, $[\vec{P}] \Lambda^{(P)}$, of the isospin-$1/2$ channel to determine the finite-volume energy levels shown in Fig.~\ref{spec_A1}. The subscripts on the ``two-meson" operators refer to the relative momentum types. The $^*$ indicates operators that were not used in the $16^{3}$ determination of the $[000]A_1^+$ spectrum. The number of $\bar{q}q$ operators used, $\mathfrak{n}$, is indicated by $(\singleop) \times~\mathfrak{n}$, where $\mathbf{\Gamma}$ represents some combination of Dirac $\gamma$-matrices and up to three (two) spatial covariant derivatives at rest (non-zero momentum) -- see Section~\ref{subsec:spectral_extraction} for further details.}
\label{tab_ops_A1}
\end{table}

\begin{table}
\begin{center}
\begin{tabular}{c|c|c|c|c}
$[000]T_1^-$                    & $[001]E_2$                      & $[011]B_1$                 & $[011]B_2$                & $[111]E_2$ \\
\hline
\hline
$D_{001}\pi_{00\sm1}$           & $D_{011}\pi_{00\sm1}$           & $D_{001}\pi_{0\sm10}$    & $D_{111}\pi_{00\sm1}$       & $D_{011}\pi_{\sm100}$ \\
$D_{011}\pi_{0\sm1\sm1}$        & $D_{001}\pi_{0\sm1\sm1}$        &$D_{011}\pi_{\sm1\sm10}$   &$D_{011}\pi_{\sm1\sm10}$    &$D_{001}\pi_{\sm1\sm10}$\\
$D_{111}\pi_{\sm1\sm1\sm1}$    &$D_{111}\pi_{0\sm1\sm1}$       &$D_{011}\pi_{00\sm2}$     &$D_{001}\pi_{\sm1\sm1\sm1}$  &$D_{002}\pi_{\sm1\sm1\sm1}$\\
$D_{001}\eta_{00\sm1}$          & $D_{011}\pi_{\sm1\sm1\sm1}$     & $D_{001}\eta_{0\sm10}$     & $D_{111}\eta_{00\sm1}$    & $D_{011}\eta_{\sm100}$ \\
$D_{011}\eta_{0\sm1\sm1}$       & $D_{011}\eta_{00\sm1}$          & $D_{011}\eta_{\sm1\sm10}$  &$D_{011}\eta_{\sm1\sm10}$ &$D_{001}\eta_{\sm1\sm10}$\\
$^*D_{111}\eta_{\sm1\sm1\sm1}$&$D_{001}\eta_{0\sm1\sm1}$   &$D_{s\,001}\bar{K}_{0\sm10}$ &$D_{s\,111}\bar{K}_{00\sm1}$ &$D_{s\,011}\bar{K}_{\sm100}$\\
$D_{s\,001}\bar{K}_{00\sm1}$&$D_{111}\eta_{0\sm1\sm1}$&$D_{s\,011}\bar{K}_{\sm1\sm10}$&$D_{s\,011}\bar{K}_{\sm1\sm10}$&$D_{s\,001}\bar{K}_{\sm1\sm10}$\\
$D_{s\,011}\bar{K}_{0\sm1\sm1}$ & $D_{s\,011}\bar{K}_{00\sm1}$    &                                 &                                 & \\
                                & $D_{s\,001}\bar{K}_{0\sm1\sm1}$ &                                 &                                 & \\[1ex]

$(\singleop) \times 16$         & $(\singleop) \times 18$         & $(\singleop) \times 28$      & $(\singleop) \times 32$  &$(\singleop) \times 42$\\
\hline
\end{tabular}
\end{center}
\caption{As Table \ref{tab_ops_A1} but for Fig.~\ref{spec_P}. The $^*$ indicates operators that were not used in our determination of the $T^{-}_{1}$ spectrum on the $24^{3}$ volume.}
\label{tab_ops_P}
\end{table}

\begin{table}
\begin{center}
\begin{tabular}{c|c|c|c}
$[000]E^+$                      & $[000]T_2^+$                     & $[001]B_1$                        & $[001]B_2$                          \\
\hline
\hline
$D_{001}\pi_{00\sm1}$           & $D_{011}\pi_{0\sm1\sm1}$         & $D_{011}\pi_{00\sm1}$             & $D_{111}\pi_{0\sm1\sm1}$            \\
$D_{011}\pi_{0\sm1\sm1}$        & $D_{111}\pi_{\sm1\sm1\sm1}$      & $D_{001}\pi_{0\sm1\sm1}$          & $D_{011}\pi_{\sm1\sm1\sm1}$         \\
$D_{002}\pi_{00\sm2}$           & $D_{011}\eta_{0\sm1\sm1}$        & $D_{011}\eta_{00\sm1}$            & $D_{111}\eta_{0\sm1\sm1}$           \\
$D_{001}\eta_{00\sm1}$          & $D_{111}\eta_{\sm1\sm1\sm1}$     & $D_{001}\eta_{0\sm1\sm1}$         & $D_{011}\eta_{\sm1\sm1\sm1}$        \\
$D_{011}\eta_{0\sm1\sm1}$       & $D_{s\,011}\bar{K}_{0\sm1\sm1}$  & $D_{s\,011}\bar{K}_{00\sm1}$      & $D_{s\,111}\bar{K}_{0\sm1\sm1}$     \\
$D_{s\,001}\bar{K}_{00\sm1}$    &                                  & $D_{s\,001}\bar{K}_{0\sm1\sm1}$   & $D_{s\,011}\bar{K}_{\sm1\sm1\sm1}$  \\
$D_{s\,011}\bar{K}_{0\sm1\sm1}$ &                                  &                                   &                                     \\[1ex]
 
$(\singleop) \times 12$         & $(\singleop) \times 15$          & $(\singleop) \times 16$           & $(\singleop) \times 16$             \\
\hline
\end{tabular}
\end{center}
\caption{As Table \ref{tab_ops_A1} but for Fig.~\ref{spec_D}.}
\label{tab_ops_D}
\end{table}

\begin{table}
\begin{center}
\begin{tabular}{c|c|c|c}
\multicolumn{1}{c|}{$[000]A_1^+$}     &\multicolumn{1}{c|}{$[001]A_1$} & \multicolumn{1}{c|}{$[011]A_1$} & \multicolumn{1}{c}{$[111]A_1$} \\
\hline \hline
$D_{000}\pi_{000}$                    & $D_{001}\pi_{000}$             & $D_{011}\pi_{000}$              & $D_{111}\pi_{000}$          \\  
$D_{001}\pi_{00\sm1}$                 & $D_{000}\pi_{00\sm1}$          & $D_{000}\pi_{0\sm1\sm1}$        & $D_{000}\pi_{\sm1\sm1\sm1}$ \\
$D_{011}\pi_{0\sm1\sm1}$              & $D_{011}\pi_{00\sm1}$          & $D_{001}\pi_{0\sm10}$           & $D_{011}\pi_{\sm100}$       \\
$D_{111}\pi_{\sm1\sm1\sm2}$           & $D_{001}\pi_{0\sm1\sm1}$       & $D_{111}\pi_{00\sm1}$           & $D_{001}\pi_{\sm1\sm0}$     \\
                                      & $D_{111}\pi_{0\sm1\sm1}$       & $D_{001}\pi_{\sm1\sm1\sm1}$     & $D_{002}\pi_{\sm1\sm1\sm1}$ \\
                                      & $D_{011}\pi_{\sm1\sm1\sm1}$    & $D_{011}\pi_{\sm1\sm10}$        & $D_{111}\pi_{00\sm2}$       \\
                                      & $D_{002}\pi_{00\sm1}$          & $D_{002}\pi_{0\sm1\sm1}$        &                             \\
                                      & $D_{001}\pi_{00\sm2}$          & $D_{011}\pi_{00\sm2}$           &                             \\
\hline
\end{tabular}
\end{center}
\caption{The interpolating operators used in the isospin-$3/2$ channel to determine the finite-volume energy levels in Fig.~\ref{fig:I32_A1_spectra}.}
\label{tab_I32_ops_A1}
\end{table}

\begin{table}
\begin{center}
\begin{tabular}{c|c|c|c|c}
$[000]T_1^-$                    & $[001]E_2$                      & $[011]B_1$                 & $[011]B_2$                  & $[111]E_2$                 \\
\hline
\hline
$D_{001}\pi_{00\sm1}$           & $D_{011}\pi_{00\sm1}$           & $D_{001}\pi_{0\sm10}$      & $D_{111}\pi_{00\sm1}$       & $D_{011}\pi_{\sm100}$       \\
$D_{011}\pi_{0\sm1\sm1}$        & $D_{001}\pi_{0\sm1\sm1}$        & $D_{011}\pi_{\sm1\sm10}$   & $D_{001}\pi_{\sm1\sm1\sm1}$ & $D_{001}\pi_{\sm1\sm10}$    \\
$D_{111}\pi_{\sm1\sm1\sm1}$     & $D_{111}\pi_{0\sm1\sm1}$        & $D_{002}\pi_{0\sm1\sm1}$   & $D_{011}\pi_{\sm1\sm10}$    & $D_{002}\pi_{\sm1\sm1\sm1}$ \\
$D_{002}\pi_{00\sm2}$           & $D_{011}\pi_{\sm1\sm1\sm1}$     & $D_{011}\pi_{00\sm2}$      &                             & $D_{111}\pi_{00\sm2}$       \\
\hline
\end{tabular}
\end{center}
\caption{As Table \ref{tab_I32_ops_A1} but for Fig.~\ref{fig:I32_nonA1_spectra}. }
\label{tab_I32_ops_P}
\end{table}

\section{Parametrisation Variations}
\label{app:parametrisation}

In Table~\ref{tab_elastic_variations} we show the elastic $S$ and $P$-wave parametrisations used in Section~\ref{sec:swave_elastic_param_vars}. In Tables~\ref{tab_coupled_variations} and \ref{tab_D_wave_par_vars} we show the coupled-channel $S$ and $D$-wave $t$-matrix parametrisations used in Sections~\ref{sec:coupled_scat_swave} and~\ref{sec:Dwave_param_vars} respectively. 

\begin{table}
\begin{center}
\begin{tabular}{lccc}
Parametrisation                                          & $N_\mathrm{pars}^{(\ell=0)}$ &  $N_\mathrm{pars}^{(\ell=1)}$ & $\chi^2/N_\mathrm{dof}$\\
\hline
\multicolumn{4}{l}{K-matrix with Chew-Mandelstam $I(s)$ \& $K_1=\frac{g_1^2}{m_1^2-s}+\gamma_1$}\\
\quad\quad\, $K=\frac{g^2}{m^2-s}$                              & 2 & 3 & {\it 2.73}\\ 
(a) \quad $K=\frac{g^2}{m^2-s}+\gamma^{(0)}$                 & 3 & 3 & 1.64\\ 
(b) \quad $K=\frac{g^2}{m^2-s}+\gamma^{(1)} s$               & 3 & 3 & 1.63\\
(c) \quad $K=\frac{(g^{(1)})^2 s}{m^2-s} + \gamma^{(0)}$     & 3 & 3 & 1.64\\
(d) \quad $K=\frac{(g+g^{(1)})^2 s}{m^2-s}$                & 3 & 3 & 1.66\\[0.3ex]
\hline
\multicolumn{4}{l}{K-matrix with Chew-Mandelstam $I(s)$ \& $K_1=\frac{g_1^2}{m_1^2-s}$}\\
\quad\quad\, $K=\frac{g^2}{m^2-s}$                              & 2 & 2 & {\it 2.94}\\ 
(e) \quad $K=\frac{g^2}{m^2-s}+\gamma^{(0)}$                 & 3 & 2 & 1.82\\ 
(f) \quad $K=\frac{g^2}{m^2-s}+\gamma^{(1)} s$               & 3 & 2 & 1.82\\
\hline
\multicolumn{4}{l}{K-matrix with $I(s)=-i\rho(s)$ \& $K_1=\frac{g_1^2}{m_1^2-s}+\gamma_1$}\\
\quad\quad\, $K=\frac{g^2}{m^2-s}$                              & 2 & 3 & {\it 2.72}\\ 
(g) \quad $K=\frac{g^2}{m^2-s}+\gamma^{(0)}$                 & 3 & 3 & 1.61\\ 
(h) \quad $K=\frac{g^2}{m^2-s}+\gamma^{(1)} s$               & 3 & 3 & 1.64\\[0.3ex]
\hline
\multicolumn{4}{l}{K-matrix with $I(s)=-i\rho(s)$ \& $K_1=\frac{g_1^2}{m_1^2-s}$}\\
\quad\quad\, $K=\frac{g^2}{m^2-s}$                              & 2 & 2 & {\it 2.93}\\ 
(i) \quad $K=\frac{g^2}{m^2-s}+\gamma^{(0)}$                 & 3 & 2 & 1.81\\ 
(j) \quad $K=\frac{g^2}{m^2-s}+\gamma^{(1)} s$               & 3 & 2 & 1.80\\[0.3ex]
\hline
\multicolumn{4}{l}{Effective range expansion in $\ell=0$ \& $K_1=\frac{g_1^2}{m_1^2-s}+\gamma_1$}\\
\quad\quad\, $k_{D\pi}\cot\delta_{D\pi} = \frac{1}{a}+\frac{1}{2}r^2 k_{D\pi}^2$                   & 2 & 3 & {\it 2.68}\\
(k) \quad $k_{D\pi}\cot\delta_{D\pi} = \frac{1}{a}+\frac{1}{2}r^2 k_{D\pi}^2 + P_2 k^4_{D\pi}$  & 3 & 3 & {\it 1.91}\\[0.3ex]
\hline
\multicolumn{4}{l}{Effective range expansion in $\ell=0$ \& $K_1=\frac{g_1^2}{m_1^2-s}$}\\
\quad\quad\, $k_{D\pi}\cot\delta_{D\pi} = \frac{1}{a}+\frac{1}{2}r^2 k_{D\pi}^2$                   & 2 & 2 & {\it 2.90}\\
\quad\quad\, $k_{D\pi}\cot\delta_{D\pi} = \frac{1}{a}+\frac{1}{2}r^2 k_{D\pi}^2 + P_2 k^4_{D\pi}$  & 3 & 2 & {\it 2.09}\\[0.3ex]
\hline
Breit-Wigner \quad $t=\frac{1}{\rho}\frac{m\Gamma}{m^2-s-im\Gamma}$ \, \& \\
\quad\quad\, $K_1=\frac{g_1^2}{m_1^2-s}$ & 2 & 2 & {\it 2.93}\\
\quad\quad\, $K_1=\frac{g_1^2}{m_1^2-s}+\gamma_1$  & 2 & 3 & {\it 2.72}\\[0.3ex]
\hline
\end{tabular}
\caption{A selection of the $S$ and $P$-wave parametrisations used for elastic $D\pi$ scattering in the isospin-$1/2$ channel in Section~\ref{sec:swave_elastic_param_vars}. $N^{(\ell)}_\mathrm{pars}$ indicates the number of free parameters in each partial wave $\ell$. $\chi^2/N_\mathrm{dof} > 1.9$ are shown in italics and these parametrisations are not included in Figs.~\ref{fig_elastic_S_kcot_amp_par_var} and \ref{fig_elastic_S_pole}.}
\label{tab_elastic_variations}
\end{center}
\end{table}

\begin{table}
\begin{center}
\scriptsize
\begin{tabular}{c|ccc|ccc|cccccc|cccccc|cc}
\multicolumn{19}{l|}{Parameters} & \multirow{3}{*}{$N_\mathrm{pars}$} & \multirow{3}{*}{$\chi^2/N_\mathrm{dof}$} \\
\multirow{2}{*}{$m$} & \multicolumn{3}{c|}{$g_i^{(0)}$} & \multicolumn{3}{c|}{$g_i^{(1)}$} & \multicolumn{6}{c|}{$\gamma_{ij}^{(0)}$} & \multicolumn{6}{c|}{$\gamma_{ij}^{(1)}$}         \\
                     & 1 & 2 & 3 & 1 & 2 & 3            & 11 & 12 & 13 & 22 & 23 & 33 & 11 & 12 & 13 & 22 & 23 & 33 \\
\hline
\hline
\tx & \tx & - & \tx & - & - & - & \tx & -   & -  & \tx & -  & \tx & - & - & - & - & - & - & 6 & \it 3.35  \\
\tx & \tx & - & \tx & - & - & - & \tx & \tx & -  & \tx & -  & \tx & - & - & - & - & - & - & 7 & \it 2.70  \\
\tx & \tx & - & \tx & - & - & - & \tx & -   &\tx & \tx & -  & \tx & - & - & - & - & - & - & 7 & \it 3.14  \\
\tx & \tx & - & \tx & - & - & - & \tx & -   & -  & \tx &\tx & \tx & - & - & - & - & - & - & 7 & \it 2.13  \\
\hline
\tx & \tx & \tx & - & - & - & - & \tx & -   & -  & \tx & -   & \tx  & - & - & - & - & - & -  & 6 & \it 13.1  \\
\tx & \tx & \tx & - & - & - & - & \tx & \tx & -  & \tx & -   & \tx  & - & - & - & - & - & -  & 7 & \it 11.7  \\
\tx & \tx & \tx & - & - & - & - & \tx & -   &\tx & \tx & -   & \tx  & - & - & - & - & - & -  & 7 & \it 2.07  \\
\tx & \tx & \tx & - & - & - & - & \tx & -   & -  & \tx & \tx & \tx  & - & - & - & - & - & -  & 7 & \it 2.07  \\
\hline
\tx & \tx & \tx & \tx & - & - & - & \tx & -   & -   & \tx & - & \tx & - & - & - & - & - & - & 7 & 1.76  \\
\tx & \tx & \tx & \tx & - & - & - & \tx & \tx & -   & \tx & - & \tx & - & - & - & - & - & - & 8 & 1.71  \\
\tx & \tx & \tx & \tx & - & - & - & \tx & \tx & \tx & \tx & - & \tx & - & - & - & - & - & - & 9 & 1.76  \\
\hline
\tx & \tx & \tx & \tx & - & - & - & - & - & - & - & - & - & \tx & -   & - & -   & - & -   & 5 & \it 2.15  \\
\tx & \tx & \tx & \tx & - & - & - & - & - & - & - & - & - & \tx & -   & - & \tx & - & -   & 6 & 1.78  \\
\tx & \tx & \tx & \tx & - & - & - & - & - & - & - & - & - & \tx & -   & - & \tx & - & \tx & 7 & 1.71  \\
\hline
\tx & \tx & \tx & \tx & \tx & -   & -   & \tx & -   & - & \tx & - & -   & - & - & - & - & - & - & 8 & 1.68  \\
\tx & \tx & \tx & \tx & \tx & -   & -   & \tx & -   & - & -   & - & \tx & - & - & - & - & - & - & 7 & \it 2.01  \\
\tx & \tx & \tx & \tx & \tx & -   & -   & \tx & -   & - & \tx & - & \tx & - & - & - & - & - & - & 8 & 1.63  \\
\tx & \tx & \tx & \tx & \tx & \tx & -   & \tx & -   & - & \tx & - & \tx & - & - & - & - & - & - & 9 & 1.66  \\
\tx & \tx & \tx & \tx & \tx & -   & \tx & \tx & -   & - & \tx & - & \tx & - & - & - & - & - & - & 9 & 1.68  \\
\hline
\end{tabular}
\caption{The $S$-wave $t$-matrix parametrisations used in Section~\ref{sec:coupled_scat_swave} where ``$\tx$'' denotes a free parameter and ``-'' a parameter fixed to zero. The channel labels are ordered by increasing mass, $1=D\pi$, $2=D\eta$ and $3=D_{s}\bar{K}$. The forms shown also included a free $P$-wave part contributing an additional 3 parameters. Forms with $\chi^2/N_\mathrm{dof} > 1.9$ (shown in italics) were not used in our final analysis as described in the text of Section~\ref{sec:coupled_scat_swave}.}
\label{tab_coupled_variations}
\end{center}
\end{table}

\begin{table}
\begin{center}
\scriptsize
\begin{tabular}{c|ccc|cccccc|cccccc|ccc}
\multicolumn{16}{l|}{Parameters} & \multirow{3}{*}{$\mathrm{Re}I(s)$} & \multirow{3}{*}{$N_\mathrm{pars}$} & \multirow{3}{*}{$\chi^2/N_\mathrm{dof}$} \\
\multirow{2}{*}{$m$} & \multicolumn{3}{c|}{$g_i^{(0)}$} & \multicolumn{6}{c|}{$\gamma_{ij}^{(0)}$} & \multicolumn{6}{c|}{$\gamma_{ij}^{(1)}$}         \\
                                         & 1 & 2 & 3            & 11 & 12 & 13 & 22 & 23 & 33 & 11 & 12 & 13 & 22 & 23 & 33 \\
\hline
\hline
\tx &  \tx & -   & -   & \tx & -   & -   & \tx & -  & \tx & - & - & - & - & - & - & CM & 5  & 1.36  \\
\tx &  \tx & -   &   - & \tx & \tx & \tx & \tx &\tx & \tx & - & - & - & - & - & - & CM & 8  & 1.28  \\
\tx &  \tx & \tx & \tx & \tx & -   & -   & \tx & -  & \tx & - & - & - & - & - & - & CM & 7  & 1.21  \\
\hline
\tx &  \tx & -   & -   & \tx & -   & -   & \tx & - & \tx & - & - & - & - & - & - & 0 & 5 & 1.35  \\
\tx &  \tx & \tx & \tx & \tx & -   & -   & \tx & - & \tx & - & - & - & - & - & - & 0 & 7 & 1.21  \\
\hline
\tx & \tx   & -  & -  & - & - & - & - & - & - & \tx & -   & -   & \tx & -   & \tx & CM & 5 & 1.16  \\
\tx & \tx & \tx & \tx & - & - & - & - & - & - & \tx & -   & -   & \tx & -   & \tx & CM & 7 & 1.26  \\
\hline
\end{tabular}
\caption{As Table~\ref{tab_coupled_variations} but for the $D$-wave parametrisations used in Section~\ref{sec:Dwave_param_vars}.}
\label{tab_D_wave_par_vars}
\end{center}
\end{table}

\clearpage

\bibliography{dpi_paper}
\bibliographystyle{JHEP}

\end{document}